\documentclass[11pt]{article}
\usepackage{fullpage}

\usepackage{array,longtable}
\newcolumntype{L}{>{$}l<{$}}
\newcolumntype{C}{>{$}c<{$}}
\newcolumntype{R}{>{$}r<{$}}

\usepackage{amsmath,amssymb}
\usepackage{graphicx}
\usepackage[nosort]{cite}
\usepackage{xcolor}

\usepackage{listings}

\newcommand{\somega}[2]{{\bf S[{\scriptstyle#1}]{#2}}}

\numberwithin{equation}{section}
\allowdisplaybreaks

\makeatletter
\g@addto@macro\bfseries{\boldmath}
\makeatother

\usepackage[margin=\parindent,font=small,labelfont=md,
labelsep=endash]{caption}

\newcommand{\cA}{{\mathcal{A}}\xspace}

\newcommand{\cD}{{\mathcal{D}}\xspace}

\newcommand{\cK}{{\mathcal{K}}\xspace}

\newcommand{\cM}{{\mathcal{M}}\xspace}
\newcommand{\cN}{{\mathcal{N}}\xspace}

\newcommand{\cV}{{\mathcal{V}}\xspace}

\newcommand{\bbR}{{\mathbb{R}}\xspace}

\newcommand{\bbZ}{{\mathbb{Z}}\xspace}

\newcommand{\ul}[1]{{\underline{#1}}}
\newcommand{\dd}{{\mathrm{d}}}

\def\dA{\smash[t]{\dot{A}}\vphantom{A}}
\def\dB{\smash[t]{\dot{B}}\vphantom{B}}

\usepackage{xspace} 

\def\E#1{\ensuremath{\mathrm{E}_{#1(#1)}}\xspace}
\def\e#1{\ensuremath{\mathfrak{e}_{#1(#1)}}\xspace}

\def\Mscalar{\ensuremath{\mathcal{M}_{70}}\xspace}

\newcommand{\SOom}{\ensuremath{\mathrm{SO}(8)_{\omega}}\xspace}

\makeatletter

\newcommand\SO{\@ifnextchar({\@SO}{\@@SO}}
\def\@SO(#1){\ensuremath{\mathrm{SO}(#1)}\xspace}
\def\@@SO#1(#2){\ensuremath{\mathrm{SO}^{#1}\kern-1pt(#2)}\xspace}

\newcommand\so{\@ifnextchar({\@so}{\@@so}}
\def\@so(#1){\ensuremath{\mathfrak{so}(#1)}\xspace}
\def\@@so#1(#2){\ensuremath{\mathfrak{so}^{#1}\kern-1pt(#2)}\xspace}

\newcommand\CSO{\@ifnextchar({\@CSO}{\@@CSO}}
\def\@CSO(#1){\ensuremath{\mathrm{CSO}(#1)}\xspace}
\def\@@CSO#1(#2){\ensuremath{\mathrm{CSO}^{#1}\kern-1pt(#2)}\xspace}

\newcommand\cso{\@ifnextchar({\@cso}{\@@cso}}
\def\@cso(#1){\ensuremath{\mathfrak{cso}(#1)}\xspace}
\def\@@cso#1(#2){\ensuremath{\mathfrak{cso}^{#1}\kern-1pt(#2)}\xspace}

\newcommand\SU{\@ifnextchar({\@SU}{\@@SU}}
\def\@SU(#1){\ensuremath{\mathrm{SU}(#1)}\xspace}
\def\@@SU#1(#2){\ensuremath{\mathrm{SU}^{#1}\kern-1pt(#2)}\xspace}

\newcommand\su{\@ifnextchar({\@su}{\@@su}}
\def\@su(#1){\ensuremath{\mathfrak{su}(#1)}\xspace}
\def\@@su#1(#2){\ensuremath{\mathfrak{su}^{#1}\kern-1pt(#2)}\xspace}

\newcommand\SL{\@ifnextchar({\@SL}{\@@SL}}
\def\@SL(#1){\ensuremath{\mathrm{SL}(#1)}\xspace}
\def\@@SL#1(#2){\ensuremath{\mathrm{SL}^{#1}\kern-1pt(#2)}\xspace}

\let\sl\relax
\newcommand\sl{\@ifnextchar({\@sl}{\@@sl}}
\def\@sl(#1){\ensuremath{\mathfrak{sl}(#1)}\xspace}
\def\@@sl#1(#2){\ensuremath{\mathfrak{sl}^{#1}\kern-1pt(#2)}\xspace}

\newcommand\GL{\@ifnextchar({\@GL}{\@@GL}}
\def\@GL(#1){\ensuremath{\mathrm{GL}(#1)}\xspace}
\def\@@GL#1(#2){\ensuremath{\mathrm{GL}^{#1}\kern-1pt(#2)}\xspace}

\newcommand\gl{\@ifnextchar({\@gl}{\@@gl}}
\def\@gl(#1){\ensuremath{\mathfrak{gl}(#1)}\xspace}
\def\@@gl#1(#2){\ensuremath{\mathfrak{gl}^{#1}\kern-1pt(#2)}\xspace}

\makeatother

\def\Sp(#1){\ensuremath{\mathrm{Sp}(#1)}\xspace}
\def\rmU(#1){\ensuremath{\mathrm{U}(#1)}\xspace}
\def\ISO(#1){\ensuremath{\mathrm{ISO}(#1)}\xspace}

\newcommand{\rmG}{\mathrm{G}}
\newcommand{\fg}{{\mathfrak{g}}}

\usepackage{hyperref}
\hypersetup{unicode,pdfborder={0 0 0},colorlinks=false,linktocpage=true}

\begin{document}

\titlepage
\begin{flushright}
QMUL-PH-21-05
\end{flushright}

\vspace*{0.5cm}

\begin{center}
{\bf \Large Vacua of $\omega$-deformed \SO(8) supergravity}

\vspace*{1cm} \textsc{%
David Berman$^{a\,}$\footnote{d.s.berman@qmul.ac.uk},
Thomas Fischbacher$^{b\,}$\footnote{tfish@google.com} \\
Gianluca Inverso$^{a,c\,}$\footnote{gianluca.inverso@pd.infn.it},  and 
Ben Scellier$^{b,d\,}$\footnote{benjamin.scellier@math.ethz.ch}%
} \\

\vspace*{0.5cm} %
$^a$ Centre for Theoretical Physics, School of Physical and Chemical Sciences, \\
Queen Mary University of London, 327 Mile End Road, London E1 4NS, UK\\

\vspace*{0.5cm} $^b$ 
Google Research, Brandschenkestrasse 110, 8002 Z\"urich, Switzerland\\

\vspace*{0.5cm} $^c$ 
INFN, Sezione di Padova \\
Via Marzolo 8, 35131 Padova, Italy

\vspace*{0.5cm} $^d$ 
Department of Mathematics, ETH Z\"urich,\\ 
R\"amistrasse 101, 8092 Z\"urich, Switzerland

\end{center}

\vspace*{0.5cm}

\begin{abstract}
We perform a detailed analysis of the vacua of $\omega$-deformed $\SO(8)$ supergravity in four dimensions. In particular, using \texttt{Tensorflow}-based numerical methods, we track how the equilibria of the theory change when varying the electric-magnetic deformation parameter $\omega$. Apart from describing various properties of different equilibria (390 in total), we show that as $\omega$ is deformed, the $\SO(3)$, $\cN=1$ vacuum of the de Wit-Nicolai theory becomes equivalent to a critical point in $\rmU(4)\ltimes\bbR^{12}$ gauged supergravity with a known uplift in IIB supergravity. The procedure employed here to obtain a new gauging with a guaranteed equilibrium is generic and allows one to obtain further admissible noncompact gaugings via $\omega$-deformation, all of which have guaranteed critical points, and some of which may be novel upliftable solutions.
\end{abstract}

\vspace*{0.5cm}

\newpage
\tableofcontents

\section{Introduction}\label{sec:intro}

The classification of gauged supergravities and of their vacuum solutions is notoriously an important, though extremely difficult task.
In particular, maximal supergravities with compact \SO(n) gauge groups are known to arise from sphere compactifications of ten- and eleven-dimensional supergravities \cite{deWit:1986oxb,Nastase:1999kf,Lee:2014mla,Hohm:2014qga} and their (anti-de Sitter) vacua are mapped to dual conformal field theory models by the holographic correspondence. 
Classifying such solutions requires extremisation of a complicated non-polynomial potential function, dependent on a large number of scalar field expectation values.
Given the significant technical effort required to analytically scan for solutions of the vacuum equations of one given gauged model, efficient methods for the numerical extremisation of the scalar potential become highly desirable.
Indeed, the use of \texttt{Tensorflow} and its autodifferentiation capabilities has proven extremely effective in charting the landscape of vacua of such gauged supergravities, revealing a cornucopia of novel solutions \cite{Comsa:2019rcz,Krishnan:2020sfg,Bobev:2020ttg,Bobev:2020qev} (see also \cite{Fischbacher:2008zu,Fischbacher:2009cj,Fischbacher:2010ki,Fischbacher:2011jx} for previous numerical work).

The set of interesting gauged maximal supergravities is known to be much larger than just the standard \SO(n) models arising from $S^{n-1}$ compactifications.
In the last decade especially, there has been significant progress in the classification of gauged maximal supergravities and of their higher dimensional origins, thanks to the embedding tensor formalism and to the construction of exceptional field theories (see \cite{Samtleben:2008pe,Gallerati:2016oyo,Trigiante:2016mnt} for reviews of the former, \cite{Berman:2020tqn} for the latter, and references therein).
In particular, it has become clear that some gauged models come in (continuous or discrete) families of similar, but physically inequivalent theories \cite{DallAgata:2012mfj,DallAgata:2014tph}.
Among such families one finds new models exhibiting interesting uplifts to string theory \cite{Guarino:2015jca,Guarino:2015vca,Inverso:2016eet,Malek:2017cle,Inverso:2017lrz}, as well as a landscape of new vacuum solutions one may therefore wish to study.
Furthermore, it was found that different families, discerned by their gauge groups, may be connected to each other through singular limits along both their scalar field and parameter spaces \cite{Borghese:2012qm,DallAgata:2012plb,Borghese:2013dja,Catino:2013ppa,Gallerati:2014xra}.
It becomes therefore highly desirable to learn how to extend the (\texttt{Tensorflow}-based) numerical extremisation techniques developed so far, to enable a unified analysis of such families of theories in order to study how their physical properties and landscape of solutions change while we vary their parameters, as well as to investigate the singular limits connecting different gauge groups and their solutions.
This is the main purpose of this paper.

The numerical techniques developed here are of wide application, but the focus of our analysis will be the extension of the numerical techniques described in \cite{Comsa:2019rcz} to the one-parameter family of $\omega$-deformed \SO(8) gauged maximal supergravities in four dimensions.
Such theories differ from the original de Wit--Nicolai model \cite{deWit:1982bul} in the choice of gauge connection, which is rotated by an electric-magnetic duality angle with respect to the standard choice.
While it has been proven that they do not admit an uplift to ten- or eleven-dimensional supergravity, requiring instead a violation of the section condition to in order to be embedded in exceptional field theory \cite{Lee:2015xga}, they have remained an interesting playground for several reasons.
As suggested by their exceptional field theory embedding, it may be possible that some non-standard string theory origin for such gaugings in fact exists, and studying their vacuum structure and how it changes with $\omega$ can help us reveal it.
Indeed, the \SOom models exhibit all the same ``main'', fully supersymmetric AdS$_4$ vacuum also present in the original, as well as a somewhat richer structure of less symmetric solutions compared to the $\omega=0$ theory.
Furthermore, following vacua through singular limits in $\omega$-space may connect them to other models, such as the ``dyonic CSO'' gaugings of \cite{DallAgata:2011aa}, which do admit supergravity uplift to ten or eleven dimensions \cite{Inverso:2016eet}.

As a first step, we perform a scan for vacua at a fixed value of the deformation parameter, finding 390 vacua. 
We also analyse their properties such as vaule of the cosmological constant, mass spectra, preserved gauge symmetries and superysmmetries. 
Among these, we find several new supersymmetric as well as non-suprsymmetric but perturbatively stable solutions.
These results are summarised in table~\ref{summarytab} and the following sections, while details for all extrema are given in the appendix (of the arXiv version of this paper).

Our next step is to develop a method to numerically follow a chosen solution as we vary $\omega$, reconstructing its trajectory.
Some trajectories are rather simple: the position and cosmological constant of some extrema may not depend on $\omega$, while others move in field space while preserving most of their properties such as their spectra (see for instance \cite{Borghese:2012qm,Borghese:2012zs}). 
However, we also uncover some unexpected behaviour which has not been noticed before.
Performing $\omega$-deformations of various equilibria, 
one finds that $\omega$-trajectories of
equilibria with rather different physical properties, even different
dimension of the unbroken gauge group, can merge and branch
again. This gives rise to a rather intricate web of connections
between different vacua. 
The analysis of the full web is beyond the scope
of this work and will be presented in a follow-up
publication~\cite{TOAPPEAR}.

As anticipated, sometimes such trajectories diverge when reaching special values of $\omega$, namely, some of the vevs reach infinite distance in scalar field space.
In such cases, the vacuum can be recovered as a solution of a different gauged supergravity, as done in \cite{DallAgata:2012plb,Catino:2013ppa,Borghese:2013dja,Gallerati:2014xra}, by absorbing the diverging vevs into the gauge couplings (through an \E7 duality transformation) and performing an overall rescaling to keep the latter finite in the limit.
We therefore develop techniques to peform such a procedure numerically, minimising noise and identifying such ``boundary gaugings'' and the associated solution with high confidence.
We use such tools to unveil a connection through $\omega$-space between the \SO(3), $\cN=1$ invariant solution of the original de Wit--Nicolai theory and a vacuum with similar properties in a $\rmU(4)\ltimes\bbR^{12}$ gauging that is part of the ``dyonic CSO'' class.
The analytic expression of such solution, as well as its type IIB uplift and many of its properties were recently identified by three of the authors in \cite{Berman:2021ynm}, using the numerical results of this paper as starting point.

This paper is organised as follows.
In the next section we summarise the basic structure of gauged maximal supergravities in four dimensions and in particular of the \SOom models.
In section~\ref{sec:pi8vacua} we look for vacua of the $\omega=\pi/8$ theory. We give a summary of the 390 solutions we find and describe the supersymmetric and stable ones in some detail.
Section 4 is the central focus of the paper where we present the numerical analysis of vacuum trajectories in $\omega$-space. 
Finally, in section 5, we describe the numerical strategy used to link vacua in singular limits along their $\omega$ trajectories.
This will allow us to map the the $\SO(3)$, $\cN = 1$ vacuum of the de Wit-Nicolai theory to a critical point in  $\rmU(4)\ltimes\bbR^{12}$ gauged maximal supergravity.

The code associated to this paper is available as a \texttt{Google Colab} \cite{Collab}.

\section{\texorpdfstring{\SOom}{SO(8)ω} gauged maximal supergravity}\label{sec:SO8omega}

Maximal supergravity in $D=4$ dimensions exhibits a global \E7 symmetry. 
The bosonic field content is given by the Einstein frame metric $g_{\mu\nu}$, 28 vector fields $A_\mu^\Lambda$ together with their magnetic duals $A_{\mu\,\Lambda}$, and 70 scalar fields parameterising the symmetric space $\Mscalar:=\E7 / (\SU(8)//\bbZ_2)$.%
The vector fields and their duals transform together under \E7 in the $\bf56$ representation and we write $A_\mu^M = (A_\mu^\Lambda,\,A_\mu{}_\Lambda)$.
The field content is completed by eight gravitini in the defining representation of \SU(8) and spin 1/2 fermions in the $\bf56$ of \SU(8).
We will denote the generators of \E7 by $t_\alpha$ with $\alpha$ the adjoint index.
The gaugings of $D=4$ maximal supergravity are described by the embedding tensor formalism \cite{deWit:2007kvg}.
They amount to promoting a subgroup $\rmG \in \E7$ to a local symmetry using $A_\mu^M$ to build the gauge connection.
The choice is encoded in a constant matrix (the embedding tensor) $\Theta_M{}^\alpha$ sitting the tensor product $\mathbf{56}\otimes\mathbf{133}$ of the fundamental and adjiont representations of \E7, such that the gauge covariant derivative is written as
\begin{equation}
\partial_\mu - g \,A_\mu^M \Theta_M{}^\alpha t_\alpha  \,.
\end{equation}
We see that $\Theta_M{}^\alpha$ determines which $t_\alpha$ generators are promoted to local symmetries, as well as which combinations of vector fields enter the gauge connection.
The embedding tensor must satisfy a set of algebraic consistency constraints, such that $\Theta_M{}^\alpha$ is in fact a highly degenerate matrix.
Supersymmetry as well as counting of the bosonic degrees of freedom require that $\Theta_M{}^\alpha$  only belongs to the $\bf912$ irrep contained within $\mathbf{56}\otimes\mathbf{133}$.
A quadratic constraint guarantees closure of the gauge algebra $\fg$ as well as gauge invariance of the embedding tensor itsef.
It reads
\begin{equation}\label{QC1}
  \Theta_M{}^\alpha t_\alpha{}_N{}^P \Theta_P^\beta
+ \Theta_M{}^\alpha \Theta_N{}^\beta f_{\alpha\beta}{}^\gamma
= 0  \,,
\end{equation}
where $t_\alpha{}_M{}^N$ are the \e7 generators in the $\bf56$ representation
and $f_{\alpha\beta}{}^\gamma$ are the structure constants of \e7:
$[t_\alpha,\,t_\beta]=f_{\alpha\beta}{}^\gamma t_\gamma$.
This constraint can be also re-written as
\begin{equation}\label{QC2}
\Omega^{MN}\Theta_M{}^\alpha \Theta_N{}^\beta = 0\,,
\end{equation}
where $\Omega^{MN}$ is the \E7 symplectic invariant.
The information encoded in $\Theta_M{}^\alpha$ is equivalently contained in the tensor
\begin{equation}
X_{MN}{}^P = \Theta_M{}^\alpha t_\alpha{}_N{}^P \,.
\end{equation}

The scalar fields of maximal supergravity can be encoded in a coset representative $\cV_M{}^\ul{N}$, with inverse $\cV_\ul{M}{}^N$, where the underlined index transforms under local \SU(8).
The supersymmetry transformations of the fermions are deformed by the gauging.
Such deformation is encoded in the so-called $T$-tensor, which is the embedding tensor dressed with the coset representative $\cV_{\ul M}{}^M$ and its inverse $\cV_{M}{}^{\ul M}$:
\begin{equation}\label{ttensor}
T_{\ul{MN}}{}^\ul{P} 
= 
\cV_{\ul M}{}^M \cV_{\ul N}{}^N X_{MN}{}^P \cV_P{}^\ul{P} \ .
\end{equation}
Under \su(8), the relevant \e7 representations decompose as
\begin{equation}\label{su8 branchings}
\mathbf{56}  \to \mathbf{28} + \overline{\mathbf{28}} \ ,\qquad
\mathbf{133} \to \mathbf{63} + \mathbf{70}  \ ,\qquad
\mathbf{912} \to \mathbf{36} + \overline{\mathbf{36}} 
                 + \mathbf{420} + \overline{\mathbf{420}}  \ .
\end{equation}
In particular, using \SU(8) fundamental indices $i,\,j,\ldots$, the $\overline{\bf28}\otimes\overline{\bf28}\otimes{\bf28}$ component of the $T$-tensor reads
\begin{equation}
T_{ij\,kl}{}^{mn} 
= 
  \delta_{[i}{}^{[m}  A_{1\,j][k} \, \delta_{l]}{}^{n]}
+ \frac12   A_{2}{}^{[m}{}_{ij[k} \, \delta_{l]}{}^{n]} \ ,
\end{equation}
where $A_{1\,ij}=A_{1\,(ij)}$, $A_2{}^i{}_{jkl}=A_2{}^i{}_{[jkl]}$ and $A_2{}^i{}_{ikl}=0$. 
They sit in the $\overline{\bf36}$ and $\bf420$ of \SU(8), respectively.
Complex conjugation corresponds to raising/lowering of all fundamental indices.

In this paper we will look for stationary points of the scalar potential of gauged maximal supergravity.
This is given by
\begin{equation}\label{scalpot}
V =  g^2\left(\frac{1}{24} A_2{}^i{}_{jkl} A_{2\,i}{}^{jkl} - \frac34 A_{1\,ij} A_1^{ij}\right)\ .
\end{equation}
The stationarity condition is given by
\begin{equation}\label{vac cond}
Q^{ijkl}_{(+)} 
=
  Q^{ijkl} 
+ \frac{1}{4!}\varepsilon^{ijklmnpq} Q_{mnpq} 
= 0   \,, \qquad
Q^{ijkl} 
= 
  \frac34 A_{2\,m}{}^{n[ij} A_{2\,n}{}^{pq]m} 
- A_1^{m[i} A_{2\,m}{}^{jkl]} \ .
\end{equation}

Residual supersymmetry of a vacuum solution is determined by vanishing of the supersymmetry variations of the fermions. 
They reduce to
\begin{equation}\label{susy cond}
2\cD_\mu \epsilon^i + g\sqrt2 A_1^{ij} \gamma_\mu \epsilon_j = 0 \,,\qquad
\epsilon^i A_{2\,i}{}^{jkl} = 0\,,
\end{equation}
where $\cD_\mu$ includes the spin connection of the background geometry, $\epsilon^i$ is a Majorana spinor (the supersymmetry parameter) and the fermion shifts are evaluated at the vacuum.

It is convenient to introduce an explicit bases for \e7 where the \sl(8,\bbR) or \su(8) subalgebras are manifest.
The latter is obtained by decomposing \e7 according to \eqref{su8 branchings}. 
The $\bf28$ can be represented as a pair of antisymmetrised indices such as $ij$, $kl$, {et cetera}, so that we can write
\begin{equation}
t_\alpha{}_{\ul M}{}^{\ul N} = 
\begin{pmatrix}
  2 \delta_{[i}{}^{[k} \lambda_{j]}{}^{l]}  &
    \sigma_{ijkl} \\[1ex]
    \sigma^{ijkl} &
- 2 \delta_{[k}{}^{[i} \lambda_{l]}{}^{j]}
\end{pmatrix} \ ,
\end{equation}
where $\lambda_{i}{}^{j}$ generate \su(8,\bbR) and $\sigma_{ijkl}$ satisfy $\sigma^{ijkl} = \frac{1}{4!}\varepsilon^{ijklmnpq} \sigma_{mnpq}$, where we remind the reader that raising/lowering of all \su(8) indices corresponds to complex conjugation.
We used underlined indices $\ul M, \ul N$ above, because in the following this basis will be associated with objects transforming under the local \SU(8).
The 70 scalar fields of maximal supergravity are identified with $\sigma_{ijkl}$ and we denote them  $\phi_{ijkl}$. 
Going from single- to double-index notation (and back), we normalise double-index sums with weight one, such that, for instance, the identity matrix in the $\bf28$ representation of \SU(8) is mapped to $\delta_{ij}^{kl}$.

Decomposing \e7 under its \sl(8,\bbR) subalgebra gives decompositions entirely analogous to \eqref{su8 branchings}.
Using indices $A,\,B,\ldots$ for the $\bf8$ of \sl(8,\bbR) and double-index notation, we then write
\begin{equation}
t_\alpha{}_M{}^N = 
\begin{pmatrix}
  2 \delta_{[A}{}^{[C} \Lambda_{B]}{}^{D]}  &
    \Sigma_{ABCD} \\[1ex]
    \frac{1}{4!}\varepsilon^{ABCDEFGH} \Sigma_{EFGH} &
- 2 \delta_{[C}{}^{[A} \Lambda_{D]}{}^{B]}
\end{pmatrix} \ ,
\end{equation}
where $\Lambda_{A}{}^{B}$ generate \sl(8,\bbR) and $\Sigma_{ABCD}$ are real and completely antisymmetric.

From now on we will take $\bf56$ indices $M,\,N,\ldots$ transforming under the gobal \E7 symmetry to be in the \sl(8) basis and underlined $\bf56$ indices $\ul M,\,\ul N,\ldots$ transforming under the local \SU(8) to be in the \su(8) basis.
The map between \sl(8,\bbR) and \su(8) bases is obtained by introducing a set of chiral $\gamma$ matrices for the common \so(8) subalgebra.
We choose conventions such that the $\bf8$ of \su(8) is identified with $\mathbf8_v$ (and we keep using indices $i,j,\ldots$ for it) and the $\bf8$ of \sl(8) with $\mathbf8_s$ (with indices $A,B,\ldots$).
Raising/lowering indices with Kronecker $\delta$s, we then use $\gamma^{ij}{}_{AB}$ to implement the triality transformation relating these two representations and construct the map
\begin{equation}
B_{\ul M}{}^{N}\ =\ 
\frac1{4\sqrt2}\begin{pmatrix}
\gamma_{ij}{}^{AB} & -i\,\gamma_{ij\,AB} \\
\gamma^{ij\,AB} & i\,\gamma^{ij}_{AB}
\end{pmatrix}\,.
\end{equation}
The coset representative appearing in \eqref{ttensor} can then be parametrised as
\begin{equation}\label{cosetrepr}
\cV = \exp\left( \frac{1}{4!}\phi_{ijkl} t^{ijkl} \right)\,\cdot\,B
\end{equation}
in terms of the generators associated to $\sigma_{ijkl}$ above, which we define as
\begin{equation}
t^{ijkl}{}_{\underline M}{}^{\underline N} = 
\begin{pmatrix}
0 & 4!\,\delta^{[ij}_{mn}\delta^{kl]}_{pq} \\
\epsilon^{ijkl\,mn\,pq}
\end{pmatrix}\,,\qquad t^{ijkl}{}_{\underline M}{}^{\underline N}\,t^{mnpq}{}_{\underline N}{}^{\underline M}=48\,\epsilon^{ijklmnpq}\,.
\end{equation}
The scalar fields $\phi_{ijkl}$ satisfy the reality condition $\phi^{ijkl} = \frac{1}{4!}\varepsilon^{ijklmnpq} \phi_{mnpq}$.

We can now introduce the embedding tensor of the \SOom models.
Decomposing the two indices of $\Theta_M{}^\alpha$ in terms of \sl(8,\bbR) indices, we have 
\begin{equation}
\Theta_M{}^\alpha = \begin{pmatrix}
\Theta_{AB\,}{}^C{}_D  & \Theta_{AB\,}{}^{CDEF} \\
\Theta^{AB\,}{}^C{}_D  & \Theta^{AB\,}{}^{CDEF}
\end{pmatrix}\,,
\end{equation}
where groups of indices are antisymmetrised, reflecting the first two branchings in \eqref{su8 branchings}.
The \SOom gaugings are defined by an embedding tensor in the ${\bf36}'+{\bf36}$ of \sl(8,\bbR), corresponding to symmetric matrices $\theta_{AB}$ and $\xi^{AB}$, following the notation in \cite{DallAgata:2011aa}.
The \SOom gaugings are defined by\footnote{To avoid any confusion on the normalisations, we state that the main vacuum of the theories, present at $\phi_{ijkl}=0$ for any $\omega$, has cosmological constant $-6g^2$.}
\begin{align}\label{SO8 embtens}
\Theta_{AB}{}^C{}_D = \delta_{[A}{}^C \theta_{B]D} \,,\quad
\Theta^{AB}{}^C{}_D = \delta_{D}{}^{[A} \xi^{B]C} \,,\qquad
\theta_{AB} = \cos\omega \ \delta_{AB}\,,\quad
\xi^{AB} = \sin\omega \ \delta^{AB}\,,
\end{align}
with $\Theta_M{}^{ABCD}=0$.
For any value of $\omega$, the gauge group is \SO(8).
However, these models are physically inequivalent for $\omega\in[0,\pi/8]$, with other values being mapped to this range by \E7 transformations \cite{DallAgata:2012mfj,DallAgata:2014tph}.
The original de Wit--Nicolai \SO(8) gauged supergravity \cite{deWit:1982bul} corresponds to $\omega=0\ \mathrm{mod}\ \pi/4$.
A much larger class of gaugings (with different gauge groups) can be defined for more general $\theta_{AB}$ and $\xi^{AB}$ \cite{DallAgata:2011aa} and we will come back to it later.

\section{Vacua of \texorpdfstring{SO(8)$_{\pi/8}$}{SO(8)π/8} gauged maximal supergravity}\label{sec:pi8vacua}

\subsection{Computational conventions and numerical extremisation}

The physical vacua, as well as the unstable equilibria in the supergravity potential usually are saddle points, characterized by~$|g^{-2}\nabla V(\phi)|=0$, where $\nabla V(\phi)$ is the gradient of the potential on the scalar manifold. If we want to interpret them as minima, we need to introduce an auxiliary function~$S$ that measures stationarity-violation. The computationally most straightforward and most generic definition here is~$S:=|g^{-2}\nabla V(\phi)|^2$, where the gradient is with respect to the parameters of the scalar function. 
This function is somewhat convenient to compute via sensitivity backpropagation, and useful for finding critical points, but physically not very meaningful. The physically more relevant gradient (which gives us proper normalization of the scalar field fluctuations'kinetic term) has to be taken with respect to local coordinates on the scalar manifold's coset space. 
For the purpose of finding equilibria, it does not matter much which variant is used, but the computational advantage of expressing the stationarity violation in the latter form, here given by \eqref{vac cond}, is that this avoids one  
layer of backpropagation through matrix exponentiation, since in this approach, we use the fermion shifts~$A_1, A_2$ anyway in order to compute the potential.
Hence, we shall take $S:=|g^{-2}Q_{(+)}|^2$.

An appropriate choice of parametrisation of the scalar manifold is important for an effective numerical search of vacua, and also to present itse results in a convenient form.
In~\cite{Comsa:2019rcz}, the scalar potential of $\SO(8)$ supergravity is
discussed in terms of a parametrization of the 56-bein as the
exponential of a $\mathfrak{e}_{7(7)}$ Lie algebra element. 
The 70=35+35 scalar fields can be organized
in the form of two symmetric traceless matrices, contained in the ${\bf
8}_s\otimes {\bf 8}_s$ and ${\bf 8}_c\otimes
{\bf 8}_c$ representations of $\so(8)$, respectively.
Denoting these matrices $M_{AB}$ and $M_{\dA\dB}$, the scalar fields $\phi_{ijkl}$ appearing in \eqref{cosetrepr} are written as 
\begin{equation}
\phi_{ijkl} = \frac14\left(
\gamma_{ijkl}{}^{AB} M_{AB} +
i\,\gamma_{ijkl}{}^{\dA\dB} M_{\dA\dB} \right) \,,
\end{equation}
where dotted indices $\dA,\,\dB,\,\ldots$ correspond to the $\mathbf8_c$ irrep of \so(8).
The results in this paper are based on a choice of \so(8) chiral $\gamma$ matrices described in equation (A.2) of \cite{Comsa:2019rcz}.
Notice that $\gamma^{ijkl\,AB}$ is self-dual in its $\mathbf8_v$ indices, while $\gamma^{ijkl\,\dA\dB}$ is anti self-dual.
The local \SO(8) invariance of the theory implies that we actually only have $70-28=42$ relevant scalar fields. 
In our search for vacua, we can generally fix this redundancy by diagonalizing one of the two symmetric traceless matrices $M_{AB}$ and $M_{\dA\dB}$. 
Furthermore, this approach usually nicely exposes the residual symmetry via eigenvalue degeneracies of a diagonal matrix and for these reasons we will present our results in terms of these two matrices.

In~\cite{Comsa:2019rcz}, most computations were performed in a~\emph{non-orthonormal} (with respect to the Lie algebra's inner product) basis of~\e7 where the basis for each of~${\bf 35}_{v,s,c}$ is given in the form of the seven matrices~$\text{diag}(+1, -1, 0, \ldots 0)$, $\text{diag}(0, +1, -1, 0, \ldots 0)$, \dots, $\text{diag}(0, \ldots, +1, -1)$, followed by the~28 symmetric off-diagonal matrices that have a single~$+1$ in their upper triangular part, by lexicographically ordered appearance. 
The main computational disadvantage of using such a non-orthonormal basis is that the matrices that represent some important symmetric bilinear(/sesquilinear) forms take on a non-symmetric form.
Unfortunately, this then makes it awkward to use existing numerical libraries to obtain orthogonal bases that are aligned with the principal axes: Generic, QR-decomposition based eigenvalue routines do not \emph{per se} return orthogonal eigenbases, so if there are degenerate eigenvalues (as usually is the case here), these tend to be of only poor numerical quality -- even more so when backpropagation is needed. Hence, the computations in this work and the accompanying code uses a basis of~\e7 in which the Killing form is (pseudo-)orthonormal, up to a multiplicative constant. This is obtained from using the lexicographic ordering on the 35 4-tuples~$(0, i, j, k),\;0<i<j<k<8$, and the gamma matrices~$\gamma^{ijkl}{}_{AB}$, $\gamma^{ijkl}{}_{\dA\dB}$.

With the conventions above in place and apart from the change of basis for the scalar fields we just described, the simplest approach to numerically extremise the scalar potential~\eqref{scalpot} for the \SOom embedding tensor~\eqref{SO8 embtens} is then to start from the Google colab notebook\footnote{This has been updated to TensorFlow2 and is now available under the url~\href{https://aihub.cloud.google.com/p/products\%2F74df893f-1ede-49c5-9f83-cfb290c05386/v/2}{https://aihub.cloud.google.com/p/products\%2F74df893f-1ede-49c5-9f83-cfb290c05386/v/2} } that was published alongside~\cite{Comsa:2019rcz} and, following Eq.~(2.20) from~\cite{deWit:2013ija}, to replace the line %
\begin{center}
\begin{tabular}{c}
\begin{lstlisting}[language=Python,basicstyle=\scriptsize]
t_uv = t_u_klKL + t_v_klIJ
\end{lstlisting}
\end{tabular}
\end{center}
with the following code that introduces a phase factor,
such as for~$\omega=\pi/8$\footnote{
The~$\omega$ in this work corresponds to~$-\omega$ in~\cite{deWit:2013ija}.}:
\begin{center}
\begin{tabular}{c}
\begin{lstlisting}[language=Python,basicstyle=\scriptsize]
omega = numpy.pi * 1j / 8
t_w = tf.math.exp(tf.constant(omega, dtype=tf.complex128))
t_uv = t_u_klKL * t_w + t_v_klIJ * tf.math.conj(t_w)
\end{lstlisting}
\end{tabular}
\end{center}
Conveniently, this small modification allows independent validation of
many of the claims made in the present article about the properties of
equilibria for specific values of~$\omega$.


\subsection{Numerical solutions at \texorpdfstring{$\omega=\pi/8$}{ω=π/8}}

All equilibria with residual supersymmetry are automatically stable, and hence also satisfy the Breitenlohner-Freedman (B.F.) bound~\cite{Breitenlohner:1982bm,Breitenlohner:1982jf}. In total, we here found 12~such solutions, listed below.
Additionally, we found six~solutions that satisfy the B.F.-bound despite having no residual supersymmetry.
Beyond that, we obtained~372 further unstable critical points whose properties are documented in detail in appendix~\ref{app:allsolutions} (of the arXiv version of this work).
An overview over these solutions is given in the table below, followed by a detailed description of the supersymmetric and non-supersymmetric perturbatively stable equilibria.%
\footnote{We denote each vacuum by ${\mathrm{\bf S}{\bf [\omega/\pi]}}$ followed by a number reflecting its cosmological constant.}

A few comments are in place. 
First, the known non-supersymmetric $\rm G_2$ invariant vacuum \cite{DallAgata:2012mfj,Borghese:2012qm} is not present in our findings. 
This is because such vacuum is extremely hard to find numerically on the full 70-dimensional scalar manifold, due to a very small basin of attraction.
Second, a non-supersymmetric, unstable vacuum with residual $\SO(3)^2$ symmetry was found in \cite{Borghese:2013dja}. 
We have found several vacua with similar propertiese but were not able to match this vacuum with one of our numerical results.

Another important comment to make is that we find several vacua that appear to have cosmological constants (and partially, mass spectra) identical to vacua found for $\omega=0$.
In particular, we have confirmed independence on the value of $\omega$ for the vacua (apart from the fully supersymmetric vacuum) $\mathrm{\bf S}\mathbf{0847213}$, $\mathrm{\bf S}\mathbf{1200000}$, $\mathrm{\bf S}\mathbf{1800000}$, $\mathrm{\bf S}\mathbf{2279257}$ and $\mathrm{\bf S}\mathbf{2279859}$ also found in \cite{Comsa:2019rcz} for $\omega=0$.
The most likely explanation for this observation is that these vacua can be obtained within some consistent truncation in which the $\omega$ deformation is trivialised, meaning that $\omega$ can be reabsorbed into a redefinition of the fields present in the truncation.
A clear instance of this situation is the $\rmU(1)^2$ invariant vacuum $\mathrm{\bf S}\mathbf{1200000}$, which can be identified within the $\cN=6$ truncation of the theory. 
It is known that $\omega$ is trivial within this truncation \cite{Inverso:2015viq} and therefore it should be extected that such vacuum can be found for any $\omega$ with the same cosmological constant.\footnote{%
Mass spectra within the $\cN=6$ theory will also be $\omega$-independent, while those of truncated fields can change.}

\small
\begin{longtable}{|L|R|L|C|C|C|C|}
\caption{Summary of vacua at $\omega=\pi/8$ (continued).}\\
\hline
\text{Name}&V/g^2&|\nabla V/g^2|&\text{Symmetry}&\text{SUSY}&\text{Stable}&\text{Ref.}\\
\hline\hline
\endhead
\caption{Summary of vacua at $\omega=\pi/8$.}\\
\hline
\text{Name}&V/g^2&|\nabla V/g^2|&\text{Symmetry}&\text{SUSY}&\text{Stable}&\text{Ref.}\\
\hline\hline
\endfirsthead
\hline
\endfoot
{\bf S[{\scriptstyle 1/8}]00674824}&-6.7482420420&<{2.5}\cdot10^{-11}&\mathfrak{so}(7)&&&\text{\cite{DallAgata:2012mfj,Borghese:2012qm}}\\
{\bf S[{\scriptstyle 1/8}]00703973}&-7.0397313061&<{9.4}\cdot10^{-12}&\mathfrak{g}_2&\mathcal{N}={1}&\checkmark&\text{\cite{DallAgata:2012mfj,Borghese:2012qm}}\\
{\bf S[{\scriptstyle 1/8}]00777052}&-7.7705274361&<{1.7}\cdot10^{-14}&\mathfrak{so}(7)&&&\text{\cite{DallAgata:2011aa,DallAgata:2012mfj}}\\
{\bf S[{\scriptstyle 1/8}]00794300}&-7.9430084586&<{5.4}\cdot10^{-12}&\mathfrak{g}_2&\mathcal{N}={1}&\checkmark&\text{\cite{DallAgata:2012mfj,Borghese:2012qm}}\\
{\bf S[{\scriptstyle 1/8}]00835418}&-8.3541893128&<{4.9}\cdot10^{-11}&\mathfrak{su}(3)+\mathfrak{u}(1)&\mathcal{N}={2}&\checkmark&\text{\cite{Borghese:2012zs}}\\
\hline
{\bf S[{\scriptstyle 1/8}]00847213}&-8.4721359550&<{6.7}\cdot10^{-12}&\mathfrak{so}(3)+\mathfrak{u}(1)+\mathfrak{u}(1)&&&\\
{\bf S[{\scriptstyle 1/8}]00858071}&-8.5807134547&<{1.6}\cdot10^{-6}&\mathfrak{su}(4)&&&\text{\cite{DallAgata:2011aa,Borghese:2012zs}}\\
{\bf S[{\scriptstyle 1/8}]00887496}&-8.8749677339&<{4.8}\cdot10^{-11}&\mathfrak{so}(3)+\mathfrak{u}(1)&&&\\
\end{longtable}

\subsubsection{Supersymmetric equilibria}

The detailed properties of solutions with residual supersymmetry are described below.

\label{S:C00600000}{\small\begin{longtable}{L}{\somega{1/8}{00600000}}\\ \mathfrak{so}(8)\rightarrow \mathfrak{so}(8),\quad \mathcal{N}={8},\quad \text{BF-stable}\\V/g^2\approx-6.0000000000,\quad |\nabla V/g^2|<{8.9}\cdot10^{-17}\\m^2/m_0^2[\psi]:\begin{minipage}[t]{10cm}\raggedright${+1.000}_{\times 8}$\end{minipage}\\m^2/m_0^2[\chi]:\begin{minipage}[t]{10cm}\raggedright${+0.000}_{\times 56}$\end{minipage}\\m^2/m_0^2[F]:\begin{minipage}[t]{10cm}\raggedright${+0.000}_{\times 28}$\end{minipage}\\m^2/m_0^2[\phi]:\begin{minipage}[t]{10cm}\raggedright${-2.000}_{\times 70}$\end{minipage}\\[6pt]{\begin{minipage}[t]{10cm}\raggedright$M_{AB}=M_{\dA\dB}=0$\end{minipage}}\vspace*{2ex}\\
\text{This is the solution at the origin with unbroken \SO(8) symmetry.}
\end{longtable}}

{\small\begin{longtable}{L}{\bf S[{\scriptstyle 1/8}]00703973}: \mathfrak{so}(8)\rightarrow \mathfrak{g}_2, \quad \mathcal{N}={1},\quad \text{BF-stable}\\V/g^2\approx-7.0397313061,\quad |\nabla V/g^2|<{9.4}\cdot10^{-12}\\m^2/m_0^2[\psi]:\begin{minipage}[t]{10cm}\raggedright\scriptsize ${+1.000}$, ${+1.500}_{\times 7}$\end{minipage}\\m^2/m_0^2[\chi]:\begin{minipage}[t]{10cm}\raggedright\scriptsize ${+0.000}_{\times 14}$, ${+0.167}_{\times 27}$, ${+1.500}_{\times 7}$, ${+6.000}_{\times 8}$\end{minipage}\\m^2/m_0^2[F]:\begin{minipage}[t]{10cm}\raggedright\scriptsize ${+0.000}_{\times 14}$, ${+0.275}_{\times 7}$, ${+2.725}_{\times 7}$\end{minipage}\\m^2/m_0^2[\phi]:\begin{minipage}[t]{10cm}\raggedright\scriptsize ${-2.242}_{\times 27}$, ${-1.425}_{\times 27}$, ${+0.000}_{\times 14}$, ${+1.551}$, ${+6.449}$\end{minipage}\\[3pt]{\begin{minipage}[t]{10cm}\raggedright$\scriptstyle M_{00}{\approx}M_{11}{\approx}M_{22}{\approx}M_{33}{\approx}M_{44}{\approx}M_{55}{\approx}M_{66}{\approx}M_{\dot0\dot0}{\approx}M_{\dot1\dot1}{\approx}M_{\dot2\dot2}{\approx}M_{\dot3\dot3}{\approx}M_{\dot4\dot4}{\approx}M_{\dot5\dot5}{\approx}M_{\dot7\dot7}\approx0.042785$, $\scriptstyle M_{77}{\approx}M_{\dot6\dot6}\approx-0.299497$\end{minipage}}\vspace*{1ex}\\
\text{This~$\rmG_2\;\mathcal{N}=1$ equilibrium was first described in~\cite{DallAgata:2012mfj,Borghese:2012qm}.}
\end{longtable}}

{\small\begin{longtable}{L}{\bf S[{\scriptstyle 1/8}]00794300}: \mathfrak{so}(8)\rightarrow \mathfrak{g}_2, \quad \mathcal{N}={1},\quad \text{BF-stable}\\V/g^2\approx-7.9430084586,\quad |\nabla V/g^2|<{5.4}\cdot10^{-12}\\m^2/m_0^2[\psi]:\begin{minipage}[t]{10cm}\raggedright\scriptsize ${+1.000}$, ${+1.500}_{\times 7}$\end{minipage}\\m^2/m_0^2[\chi]:\begin{minipage}[t]{10cm}\raggedright\scriptsize ${+0.000}_{\times 14}$, ${+0.167}_{\times 27}$, ${+1.500}_{\times 7}$, ${+6.000}_{\times 8}$\end{minipage}\\m^2/m_0^2[F]:\begin{minipage}[t]{10cm}\raggedright\scriptsize ${+0.000}_{\times 14}$, ${+0.275}_{\times 7}$, ${+2.725}_{\times 7}$\end{minipage}\\m^2/m_0^2[\phi]:\begin{minipage}[t]{10cm}\raggedright\scriptsize ${-2.242}_{\times 27}$, ${-1.425}_{\times 27}$, ${+0.000}_{\times 14}$, ${+1.551}$, ${+6.449}$\end{minipage}\\
\end{longtable}}

{\small\begin{longtable}{L}{\bf S[{\scriptstyle 1/8}]00835418}: \mathfrak{so}(8)\rightarrow \mathfrak{su}(3)+\mathfrak{u}(1), \quad \mathcal{N}={2},\quad \text{BF-stable}\\V/g^2\approx-8.3541893128,\quad |\nabla V/g^2|<{4.9}\cdot10^{-11}\\m^2/m_0^2[\psi]:\begin{minipage}[t]{10cm}\raggedright\scriptsize ${+1.000}_{\times 2}$, ${+1.778}_{\times 6}$\end{minipage}\\m^2/m_0^2[\chi]:\begin{minipage}[t]{10cm}\raggedright\scriptsize ${+0.000}_{\times 16}$, ${+0.111}_{\times 18}$, ${+1.778}_{\times 12}$, ${+2.438}_{\times 2}$, ${+6.562}_{\times 2}$, ${+7.111}_{\times 6}$\end{minipage}\\m^2/m_0^2[F]:\begin{minipage}[t]{10cm}\raggedright\scriptsize ${+0.000}_{\times 9}$, ${+0.444}_{\times 12}$, ${+3.111}_{\times 6}$, ${+4.000}$\end{minipage}\\m^2/m_0^2[\phi]:\begin{minipage}[t]{10cm}\raggedright\scriptsize ${-2.222}_{\times 12}$, ${-2.000}_{\times 16}$, ${-1.556}_{\times 18}$, ${-1.123}$, ${+0.000}_{\times 19}$, ${+2.000}_{\times 3}$, ${+7.123}$\end{minipage}\\[6 pt]{\begin{minipage}[t]{10cm}\raggedright$\scriptstyle M_{00}{\approx}M_{11}{\approx}M_{22}{\approx}M_{33}{\approx}M_{66}{\approx}M_{77}\approx-0.040301$, $\scriptstyle M_{44}\approx-0.254436$, $\scriptstyle M_{55}\approx0.496243$, $\scriptstyle M_{\dot0\dot0}{\approx}M_{\dot1\dot1}{\approx}M_{\dot2\dot2}{\approx}M_{\dot3\dot3}{\approx}M_{\dot4\dot4}{\approx}M_{\dot5\dot5}\approx0.067611$, $\scriptstyle M_{\dot6\dot6}{\approx}M_{\dot7\dot7}\approx-0.202834$\end{minipage}}\vspace*{2ex}\\
\text{This~$\SU(3)\times \rmU(1)\;\cN=2$ equilibrium was first described in~\cite{Borghese:2012zs}.}
\end{longtable}}

{\small\begin{longtable}{L}{\bf S[{\scriptstyle 1/8}]01016983}: \mathfrak{so}(8)\rightarrow \mathfrak{so}(3)+\mathfrak{so}(3), \quad \mathcal{N}={3},\quad \text{BF-stable}\\V/g^2\approx-10.1698390273,\quad |\nabla V/g^2|<{2.3}\cdot10^{-14}\\m^2/m_0^2[\psi]:\begin{minipage}[t]{10cm}\raggedright\scriptsize ${+1.000}_{\times 3}$, ${+2.250}_{\times 4}$, ${+3.000}$\end{minipage}\\m^2/m_0^2[\chi]:\begin{minipage}[t]{10cm}\raggedright\scriptsize ${+0.000}_{\times 13}$, ${+0.250}_{\times 12}$, ${+0.536}_{\times 3}$, ${+2.250}_{\times 12}$, ${+3.000}_{\times 8}$, ${+7.464}_{\times 3}$, ${+9.000}_{\times 4}$, ${+12.000}$\end{minipage}\\m^2/m_0^2[F]:\begin{minipage}[t]{10cm}\raggedright\scriptsize ${+0.000}_{\times 6}$, ${+0.750}_{\times 12}$, ${+1.268}_{\times 3}$, ${+3.750}_{\times 4}$, ${+4.732}_{\times 3}$\end{minipage}\\m^2/m_0^2[\phi]:\begin{minipage}[t]{10cm}\raggedright\scriptsize ${-2.250}_{\times 4}$, ${-2.196}$, ${-2.000}_{\times 18}$, ${-1.250}_{\times 12}$, ${-0.732}_{\times 6}$, ${+0.000}_{\times 22}$, ${+2.732}_{\times 6}$, ${+8.196}$\end{minipage}\\[6 pt]{\begin{minipage}[t]{10cm}\raggedright$\scriptstyle M_{00}\approx0.460942$, $\scriptstyle M_{11}{\approx}M_{22}{\approx}M_{33}{\approx}M_{44}\approx0.033926$, $\scriptstyle M_{55}{\approx}M_{66}{\approx}M_{77}\approx-0.198882$, $\scriptstyle M_{\dot0\dot0}{\approx}M_{\dot1\dot1}{\approx}M_{\dot2\dot2}{\approx}M_{\dot4\dot4}{\approx}M_{\dot5\dot5}{\approx}M_{\dot6\dot6}\approx-0.082478$, $\scriptstyle M_{\dot0\dot4}{\approx}M_{\dot1\dot5}{\approx}-M_{\dot2\dot6}\approx-0.116404$, $\scriptstyle M_{\dot3\dot3}{\approx}M_{\dot7\dot7}\approx0.247434$, $\scriptstyle M_{\dot3\dot7}\approx0.213508$\end{minipage}}\vspace*{2ex}\\
\text{This~$\SO(3)\times\SO(3)\;\mathcal{N}=3$ equilibrium was first described in~\cite{Gallerati:2014xra}.}
\end{longtable}}

{\small\begin{longtable}{L}{\bf S[{\scriptstyle 1/8}]01039230}: \mathfrak{so}(8)\rightarrow \mathfrak{su}(3), \quad \mathcal{N}={1},\quad \text{BF-stable}\\V/g^2\approx-10.3923048454,\quad |\nabla V/g^2|<{1.6}\cdot10^{-5}\\m^2/m_0^2[\psi]:\begin{minipage}[t]{10cm}\raggedright\scriptsize ${+1.000}$, ${+1.778}_{\times 6}$, ${+4.000}$\end{minipage}\\m^2/m_0^2[\chi]:\begin{minipage}[t]{10cm}\raggedright\scriptsize ${+0.000}_{\times 8}$, ${+0.444}_{\times 12}$, ${+1.000}_{\times 8}$, ${+1.538}_{\times 6}$, ${+1.778}_{\times 6}$, ${+4.000}$, ${+5.018}_{\times 6}$, ${+6.000}_{\times 2}$, ${+7.111}_{\times 6}$, ${+16.000}$\end{minipage}\\m^2/m_0^2[F]:\begin{minipage}[t]{10cm}\raggedright\scriptsize ${+0.000}_{\times 8}$, ${+0.444}_{\times 6}$, ${+2.000}$, ${+2.778}_{\times 6}$, ${+3.111}_{\times 6}$, ${+6.000}$\end{minipage}\\m^2/m_0^2[\phi]:\begin{minipage}[t]{10cm}\raggedright\scriptsize ${-2.222}_{\times 12}$, ${-2.000}_{\times 8}$, ${-0.889}_{\times 12}$, ${+0.000}_{\times 28}$, ${+0.778}_{\times 6}$, ${+1.551}_{\times 2}$, ${+6.449}_{\times 2}$\end{minipage}\\[6 pt]{\begin{minipage}[t]{10cm}\raggedright$\scriptstyle M_{00}{\approx}M_{11}{\approx}M_{22}{\approx}M_{55}{\approx}M_{66}{\approx}M_{77}{\approx}M_{\dot1\dot1}{\approx}M_{\dot2\dot2}{\approx}M_{\dot3\dot3}{\approx}M_{\dot4\dot4}\approx-0.048552$, $\scriptstyle M_{33}\approx0.291314$, $\scriptstyle M_{34}\approx0.378167$, $\scriptstyle M_{\dot0\dot0}\approx0.550905$, $\scriptstyle M_{\dot5\dot5}{\approx}M_{\dot6\dot6}\approx-0.048553$, $\scriptstyle M_{\dot7\dot7}\approx-0.259592$\end{minipage}}\vspace*{2ex}\\
\text{This~$\SU(3)\;\mathcal{N}=1$ equilibrium was first described in~\cite{Borghese:2012zs}.}
\end{longtable}}

\label{S:sol01200000}{\small\begin{longtable}{L}{\bf S[{\scriptstyle 1/8}]01200000}: \mathfrak{so}(8)\rightarrow \mathfrak{u}(1)+\mathfrak{u}(1), \quad \mathcal{N}={1},\quad \text{BF-stable}\\V/g^2\approx-12.0000000000,\quad |\nabla V/g^2|<{9.3}\cdot10^{-14}\\m^2/m_0^2[\psi]:\begin{minipage}[t]{10cm}\raggedright\scriptsize ${+1.000}$, ${+2.250}_{\times 4}$, ${+3.000}_{\times 2}$, ${+4.000}$\end{minipage}\\m^2/m_0^2[\chi]:\begin{minipage}[t]{10cm}\raggedright\scriptsize ${+0.000}_{\times 4}$, ${+0.007}_{\times 4}$, ${+0.048}_{\times 2}$, ${+0.250}_{\times 4}$, ${+0.360}_{\times 2}$, ${+0.536}$, ${+1.169}_{\times 4}$, ${+1.199}_{\times 2}$, ${+1.488}_{\times 2}$, ${+2.237}_{\times 2}$, ${+2.250}_{\times 4}$, ${+3.000}_{\times 2}$, ${+3.337}_{\times 2}$, ${+4.000}$, ${+4.331}_{\times 4}$, ${+6.228}_{\times 2}$, ${+7.464}$, ${+9.000}_{\times 4}$, ${+9.493}_{\times 4}$, ${+11.104}_{\times 2}$, ${+12.000}_{\times 2}$, ${+16.000}$\end{minipage}\\m^2/m_0^2[F]:\begin{minipage}[t]{10cm}\raggedright\scriptsize ${+0.000}_{\times 2}$, ${+0.088}_{\times 4}$, ${+0.268}_{\times 2}$, ${+0.750}_{\times 4}$, ${+1.268}_{\times 2}$, ${+2.000}$, ${+3.732}_{\times 2}$, ${+3.750}_{\times 4}$, ${+4.732}_{\times 2}$, ${+6.000}$, ${+6.412}_{\times 4}$\end{minipage}\\m^2/m_0^2[\phi]:\begin{minipage}[t]{10cm}\raggedright\scriptsize ${-2.250}_{\times 4}$, ${-2.240}_{\times 2}$, ${-2.196}$, ${-2.000}_{\times 4}$, ${-1.912}_{\times 4}$, ${-1.896}_{\times 2}$, ${-1.732}_{\times 2}$, ${-1.250}_{\times 4}$, ${-1.040}_{\times 2}$, ${-0.732}$, ${-0.489}_{\times 2}$, ${+0.000}_{\times 26}$, ${+0.293}_{\times 2}$, ${+1.732}_{\times 2}$, ${+2.732}$, ${+3.164}_{\times 2}$, ${+4.412}_{\times 4}$, ${+5.772}_{\times 2}$, ${+8.196}$, ${+12.436}_{\times 2}$\end{minipage}\\
\end{longtable}}

\label{S:sol01253393}{\small\begin{longtable}{L}{\bf S[{\scriptstyle 1/8}]01253393}: \mathfrak{so}(8)\rightarrow \mathfrak{so}(3), \quad \mathcal{N}={1},\quad \text{BF-stable}\\V/g^2\approx-12.5339321693,\quad |\nabla V/g^2|<{1.2}\cdot10^{-13}\\m^2/m_0^2[\psi]:\begin{minipage}[t]{10cm}\raggedright\scriptsize ${+1.000}$, ${+2.724}_{\times 3}$, ${+3.043}$, ${+3.216}_{\times 3}$\end{minipage}\\m^2/m_0^2[\chi]:\begin{minipage}[t]{10cm}\raggedright\scriptsize ${+0.000}_{\times 3}$, ${+0.016}_{\times 5}$, ${+0.043}_{\times 3}$, ${+0.052}$, ${+0.785}_{\times 5}$, ${+1.271}_{\times 5}$, ${+2.360}_{\times 5}$, ${+2.724}_{\times 3}$, ${+3.043}$, ${+3.216}_{\times 3}$, ${+4.566}_{\times 3}$, ${+4.704}_{\times 3}$, ${+7.229}$, ${+9.840}_{\times 3}$, ${+10.041}_{\times 3}$, ${+10.894}_{\times 3}$, ${+12.171}$, ${+12.864}_{\times 3}$, ${+14.020}$, ${+14.151}$\end{minipage}\\m^2/m_0^2[F]:\begin{minipage}[t]{10cm}\raggedright\scriptsize ${+0.000}_{\times 3}$, ${+0.144}_{\times 5}$, ${+1.073}_{\times 3}$, ${+1.298}$, ${+1.423}_{\times 3}$, ${+4.374}_{\times 3}$, ${+4.787}$, ${+5.009}_{\times 3}$, ${+6.703}_{\times 3}$, ${+6.872}_{\times 3}$\end{minipage}\\m^2/m_0^2[\phi]:\begin{minipage}[t]{10cm}\raggedright\scriptsize ${-2.176}$, ${-2.164}_{\times 3}$, ${-2.101}_{\times 5}$, ${-1.856}_{\times 5}$, ${-1.751}_{\times 3}$, ${-1.721}$, ${-1.176}_{\times 5}$, ${-0.329}_{\times 5}$, ${+0.000}_{\times 25}$, ${+1.897}_{\times 5}$, ${+2.540}$, ${+4.703}_{\times 3}$, ${+4.872}_{\times 3}$, ${+7.918}$, ${+8.276}$, ${+8.390}$, ${+15.764}$, ${+15.913}$\end{minipage}\\
\end{longtable}}

{\small\begin{longtable}{L}{\bf S[{\scriptstyle 1/8}]02150686}: \mathfrak{so}(8)\rightarrow \mathfrak{so}(3), \quad \mathcal{N}={1},\quad \text{BF-stable}\\V/g^2\approx-21.5068643478,\quad |\nabla V/g^2|<{2.8}\cdot10^{-13}\\m^2/m_0^2[\psi]:\begin{minipage}[t]{10cm}\raggedright\scriptsize ${+1.000}$, ${+3.488}_{\times 3}$, ${+3.616}$, ${+4.026}_{\times 3}$\end{minipage}\\m^2/m_0^2[\chi]:\begin{minipage}[t]{10cm}\raggedright\scriptsize ${+0.000}_{\times 3}$, ${+0.093}$, ${+0.210}_{\times 5}$, ${+0.616}_{\times 3}$, ${+1.041}_{\times 5}$, ${+3.488}_{\times 3}$, ${+3.616}$, ${+4.026}_{\times 3}$, ${+4.082}_{\times 5}$, ${+6.202}_{\times 3}$, ${+6.251}_{\times 3}$, ${+7.258}_{\times 5}$, ${+7.864}$, ${+12.182}_{\times 3}$, ${+12.251}_{\times 3}$, ${+13.954}_{\times 3}$, ${+14.465}$, ${+15.223}$, ${+15.424}$, ${+16.103}_{\times 3}$\end{minipage}\\m^2/m_0^2[F]:\begin{minipage}[t]{10cm}\raggedright\scriptsize ${+0.000}_{\times 3}$, ${+1.621}_{\times 3}$, ${+1.715}$, ${+2.019}_{\times 3}$, ${+2.061}_{\times 5}$, ${+5.356}_{\times 3}$, ${+5.518}$, ${+6.032}_{\times 3}$, ${+8.692}_{\times 3}$, ${+8.751}_{\times 3}$\end{minipage}\\m^2/m_0^2[\phi]:\begin{minipage}[t]{10cm}\raggedright\scriptsize ${-2.248}_{\times 5}$, ${-2.212}$, ${-2.169}_{\times 3}$, ${-1.602}$, ${-1.332}_{\times 5}$, ${-0.599}_{\times 3}$, ${+0.000}_{\times 25}$, ${+0.061}_{\times 5}$, ${+2.564}_{\times 5}$, ${+3.060}$, ${+6.692}_{\times 3}$, ${+6.751}_{\times 3}$, ${+7.952}_{\times 5}$, ${+8.669}$, ${+9.321}$, ${+9.497}$, ${+17.125}$, ${+17.351}$\end{minipage}\\
\end{longtable}}

{\small\begin{longtable}{L}{\bf S[{\scriptstyle 1/8}]02251538}: \mathfrak{so}(8)\rightarrow \emptyset, \quad \mathcal{N}={1},\quad \text{BF-stable}\\V/g^2\approx-22.5153845787,\quad |\nabla V/g^2|<{1.1}\cdot10^{-11}\\m^2/m_0^2[\psi]:\begin{minipage}[t]{10cm}\raggedright\scriptsize ${+1.000}$, ${+2.985}$, ${+3.208}$, ${+3.369}$, ${+3.605}$, ${+3.625}$, ${+5.479}$, ${+5.488}$\end{minipage}\\m^2/m_0^2[\chi]:\begin{minipage}[t]{10cm}\raggedright\scriptsize ${+0.000}$, ${+0.033}$, ${+0.034}$, ${+0.035}$, ${+0.044}$, ${+0.060}$, ${+0.568}$, ${+0.663}$, ${+0.672}$, ${+0.753}$, ${+1.020}$, ${+1.397}$, ${+1.462}$, ${+2.540}$, ${+2.770}$, ${+2.969}$, ${+2.985}$, ${+3.208}$, ${+3.292}$, ${+3.311}$, ${+3.364}$, ${+3.369}$, ${+3.605}$, ${+3.625}$, ${+4.871}$, ${+5.139}$, ${+5.479}$, ${+5.488}$, ${+5.734}$, ${+5.914}$, ${+6.545}$, ${+6.625}$, ${+6.728}$, ${+7.416}$, ${+7.515}$, ${+7.595}$, ${+7.676}$, ${+8.032}$, ${+9.407}$, ${+10.164}$, ${+10.285}$, ${+10.672}$, ${+10.993}$, ${+11.523}$, ${+11.778}$, ${+11.941}$, ${+12.830}$, ${+13.476}$, ${+13.478}$, ${+13.686}$, ${+13.998}$, ${+14.106}$, ${+14.419}$, ${+14.501}$, ${+21.918}$, ${+21.951}$\end{minipage}\\m^2/m_0^2[F]:\begin{minipage}[t]{10cm}\raggedright\scriptsize ${+0.010}$, ${+0.215}$, ${+0.253}$, ${+1.257}$, ${+1.417}$, ${+1.478}$, ${+1.492}$, ${+1.534}$, ${+1.706}$, ${+1.721}$, ${+3.139}$, ${+3.145}$, ${+4.134}$, ${+4.693}$, ${+4.713}$, ${+4.999}$, ${+5.198}$, ${+5.205}$, ${+5.503}$, ${+5.529}$, ${+7.078}$, ${+7.405}$, ${+7.820}$, ${+7.830}$, ${+8.128}$, ${+8.346}$, ${+10.257}$, ${+10.350}$\end{minipage}\\
\end{longtable}}

{\small\begin{longtable}{L}{\bf S[{\scriptstyle 1/8}]02255860}: \mathfrak{so}(8)\rightarrow \mathfrak{u}(1), \quad \mathcal{N}={1},\quad \text{BF-stable}\\V/g^2\approx-22.5586033705,\quad |\nabla V/g^2|<{1.4}\cdot10^{-11}\\m^2/m_0^2[\psi]:\begin{minipage}[t]{10cm}\raggedright\scriptsize ${+1.000}$, ${+3.457}_{\times 2}$, ${+3.521}$, ${+3.566}_{\times 2}$, ${+4.851}$, ${+5.199}$\end{minipage}\\m^2/m_0^2[\chi]:\begin{minipage}[t]{10cm}\raggedright\scriptsize ${+0.000}$, ${+0.035}_{\times 2}$, ${+0.054}_{\times 2}$, ${+0.066}$, ${+0.574}_{\times 2}$, ${+0.666}$, ${+1.168}_{\times 2}$, ${+1.518}_{\times 2}$, ${+2.917}_{\times 2}$, ${+2.936}$, ${+3.088}_{\times 2}$, ${+3.457}_{\times 2}$, ${+3.521}$, ${+3.566}_{\times 2}$, ${+3.626}$, ${+4.851}$, ${+5.199}$, ${+5.677}_{\times 2}$, ${+5.821}_{\times 2}$, ${+6.212}_{\times 2}$, ${+6.224}$, ${+6.589}$, ${+7.333}_{\times 2}$, ${+7.597}$, ${+8.435}$, ${+9.828}_{\times 2}$, ${+11.122}$, ${+11.443}_{\times 2}$, ${+11.646}_{\times 2}$, ${+12.213}$, ${+12.723}$, ${+12.952}$, ${+13.269}$, ${+13.828}_{\times 2}$, ${+14.085}$, ${+14.266}_{\times 2}$, ${+19.405}$, ${+20.797}$\end{minipage}\\m^2/m_0^2[F]:\begin{minipage}[t]{10cm}\raggedright\scriptsize ${+0.000}$, ${+0.286}_{\times 2}$, ${+1.331}_{\times 2}$, ${+1.598}_{\times 2}$, ${+1.645}$, ${+1.678}_{\times 2}$, ${+2.649}$, ${+2.919}$, ${+4.625}_{\times 2}$, ${+5.316}_{\times 2}$, ${+5.398}$, ${+5.455}_{\times 2}$, ${+5.530}$, ${+7.054}$, ${+7.479}$, ${+8.060}_{\times 2}$, ${+8.233}_{\times 2}$, ${+8.718}$, ${+9.156}$\end{minipage}\\
\end{longtable}}

{\small\begin{longtable}{L}{\bf S[{\scriptstyle 1/8}]02792705}: \mathfrak{so}(8)\rightarrow \emptyset, \quad \mathcal{N}={1},\quad \text{BF-stable}\\V/g^2\approx-27.9270575850,\quad |\nabla V/g^2|<{4.4}\cdot10^{-13}\\m^2/m_0^2[\psi]:\begin{minipage}[t]{10cm}\raggedright\scriptsize ${+1.000}$, ${+2.364}$, ${+3.384}_{\times 2}$, ${+5.625}$, ${+5.928}_{\times 2}$, ${+6.935}$\end{minipage}\\m^2/m_0^2[\chi]:\begin{minipage}[t]{10cm}\raggedright\scriptsize ${+0.014}$, ${+0.029}_{\times 2}$, ${+0.129}$, ${+0.154}$, ${+0.968}$, ${+1.370}_{\times 2}$, ${+1.846}$, ${+1.967}$, ${+2.364}$, ${+3.310}_{\times 2}$, ${+3.384}_{\times 2}$, ${+4.074}$, ${+4.236}$, ${+5.197}_{\times 2}$, ${+5.625}$, ${+5.928}_{\times 2}$, ${+6.722}$, ${+6.935}$, ${+7.212}$, ${+7.552}$, ${+7.668}$, ${+7.949}_{\times 2}$, ${+8.147}_{\times 2}$, ${+8.259}_{\times 2}$, ${+8.852}$, ${+9.110}$, ${+9.245}$, ${+9.353}$, ${+9.454}$, ${+10.756}_{\times 2}$, ${+11.915}$, ${+12.816}$, ${+12.908}$, ${+13.537}_{\times 2}$, ${+14.048}$, ${+14.233}_{\times 2}$, ${+15.006}_{\times 2}$, ${+15.018}$, ${+15.802}$, ${+22.501}$, ${+23.711}_{\times 2}$, ${+27.739}$\end{minipage}\\m^2/m_0^2[F]:\begin{minipage}[t]{10cm}\raggedright\scriptsize ${+0.199}_{\times 2}$, ${+0.488}$, ${+0.826}$, ${+1.545}_{\times 2}$, ${+3.254}$, ${+3.493}_{\times 2}$, ${+3.901}$, ${+4.301}$, ${+5.130}_{\times 2}$, ${+5.224}_{\times 2}$, ${+6.092}$, ${+6.295}$, ${+7.476}_{\times 2}$, ${+7.997}$, ${+8.363}_{\times 2}$, ${+9.315}$, ${+9.568}$, ${+10.300}$, ${+11.132}_{\times 2}$, ${+11.827}$\end{minipage}\\
\end{longtable}}

\subsubsection{Non-supersymmetric B.F.-stable equilibria}

{\small\begin{longtable}{L}{\bf S[{\scriptstyle 1/8}]01023727}: \mathfrak{so}(8)\rightarrow \mathfrak{su}(3),\quad \text{BF-stable}\\V/g^2\approx-10.2372748592,\quad |\nabla V/g^2|<{1.3}\cdot10^{-14}\\m^2/m_0^2[\psi]:\begin{minipage}[t]{10cm}\raggedright\scriptsize ${+1.261}$, ${+1.589}_{\times 6}$, ${+4.422}$\end{minipage}\\m^2/m_0^2[\chi]:\begin{minipage}[t]{10cm}\raggedright\scriptsize ${+0.184}_{\times 8}$, ${+0.539}_{\times 6}$, ${+0.580}_{\times 12}$, ${+1.238}_{\times 8}$, ${+1.486}$, ${+2.132}_{\times 6}$, ${+4.587}_{\times 6}$, ${+5.046}$, ${+5.700}$, ${+6.357}_{\times 6}$, ${+17.689}$\end{minipage}\\m^2/m_0^2[F]:\begin{minipage}[t]{10cm}\raggedright\scriptsize ${+0.000}_{\times 8}$, ${+0.128}_{\times 6}$, ${+2.321}$, ${+2.757}_{\times 6}$, ${+3.194}_{\times 6}$, ${+4.520}$\end{minipage}\\m^2/m_0^2[\phi]:\begin{minipage}[t]{10cm}\raggedright\scriptsize ${-1.641}_{\times 12}$, ${-1.504}_{\times 8}$, ${-1.275}$, ${-1.073}_{\times 6}$, ${-0.908}_{\times 12}$, ${-0.235}_{\times 8}$, ${+0.000}_{\times 20}$, ${+1.138}$, ${+5.914}$, ${+6.223}$\end{minipage}\\[6 pt]{\begin{minipage}[t]{10cm}\raggedright$\scriptstyle M_{00}\approx0.562124$, $\scriptstyle M_{11}\approx-0.078763$, $\scriptstyle M_{22}{\approx}M_{33}{\approx}M_{44}{\approx}M_{55}{\approx}M_{66}{\approx}M_{77}\approx-0.080560$, $\scriptstyle M_{\dot0\dot0}{\approx}M_{\dot1\dot1}{\approx}M_{\dot2\dot2}{\approx}M_{\dot3\dot3}{\approx}M_{\dot6\dot6}{\approx}M_{\dot7\dot7}\approx-0.059137$, $\scriptstyle M_{\dot4\dot4}\approx-0.221838$, $\scriptstyle M_{\dot5\dot5}\approx0.576658$\end{minipage}}\vspace*{2ex}\\
\text{This~$\SU(3)$ equilibrium was first described in~\cite{Borghese:2012zs}.}
\end{longtable}}

{\small\begin{longtable}{L}{\bf S[{\scriptstyle 1/8}]01024518}: \mathfrak{so}(8)\rightarrow \mathfrak{so}(3),\quad \text{BF-stable}\\V/g^2\approx-10.2451879983,\quad |\nabla V/g^2|<{8.0}\cdot10^{-14}\\m^2/m_0^2[\psi]:\begin{minipage}[t]{10cm}\raggedright\scriptsize ${+1.288}_{\times 2}$, ${+1.330}$, ${+1.723}_{\times 4}$, ${+4.360}$\end{minipage}\\m^2/m_0^2[\chi]:\begin{minipage}[t]{10cm}\raggedright\scriptsize ${+0.164}$, ${+0.234}_{\times 4}$, ${+0.274}_{\times 6}$, ${+0.363}_{\times 4}$, ${+0.371}_{\times 3}$, ${+0.645}_{\times 4}$, ${+0.738}$, ${+0.810}_{\times 3}$, ${+0.951}_{\times 2}$, ${+1.160}$, ${+1.357}$, ${+1.604}_{\times 4}$, ${+1.796}_{\times 2}$, ${+1.953}_{\times 4}$, ${+2.191}$, ${+4.075}_{\times 4}$, ${+5.152}_{\times 2}$, ${+5.322}$, ${+5.373}$, ${+5.543}_{\times 2}$, ${+6.893}_{\times 4}$, ${+17.441}$\end{minipage}\\m^2/m_0^2[F]:\begin{minipage}[t]{10cm}\raggedright\scriptsize ${+0.000}_{\times 3}$, ${+0.029}_{\times 2}$, ${+0.063}_{\times 4}$, ${+0.103}$, ${+0.111}_{\times 4}$, ${+2.398}_{\times 4}$, ${+2.481}_{\times 2}$, ${+2.701}$, ${+3.319}_{\times 4}$, ${+3.541}$, ${+4.016}_{\times 2}$\end{minipage}\\m^2/m_0^2[\phi]:\begin{minipage}[t]{10cm}\raggedright\scriptsize ${-1.713}_{\times 4}$, ${-1.665}_{\times 3}$, ${-1.584}$, ${-1.434}_{\times 6}$, ${-1.360}_{\times 6}$, ${-1.321}_{\times 5}$, ${-1.291}_{\times 3}$, ${-1.188}_{\times 2}$, ${-1.061}_{\times 4}$, ${-0.937}_{\times 4}$, ${-0.819}$, ${-0.807}$, ${-0.460}$, ${+0.000}_{\times 25}$, ${+0.163}$, ${+1.134}$, ${+5.771}$, ${+6.300}$\end{minipage}\\
\end{longtable}}

{\small\begin{longtable}{L}{\bf S[{\scriptstyle 1/8}]01027099}: \mathfrak{so}(8)\rightarrow \mathfrak{so}(3)+\mathfrak{so}(3),\quad \text{BF-stable}\\V/g^2\approx-10.2709952665,\quad |\nabla V/g^2|<{5.6}\cdot10^{-14}\\m^2/m_0^2[\psi]:\begin{minipage}[t]{10cm}\raggedright\scriptsize ${+1.155}_{\times 3}$, ${+1.872}_{\times 4}$, ${+4.108}$\end{minipage}\\m^2/m_0^2[\chi]:\begin{minipage}[t]{10cm}\raggedright\scriptsize ${+0.043}$, ${+0.177}_{\times 9}$, ${+0.236}_{\times 8}$, ${+0.456}_{\times 3}$, ${+0.733}_{\times 4}$, ${+1.346}_{\times 3}$, ${+1.667}_{\times 5}$, ${+1.761}_{\times 8}$, ${+3.653}_{\times 4}$, ${+4.620}_{\times 3}$, ${+5.992}_{\times 3}$, ${+7.489}_{\times 4}$, ${+16.433}$\end{minipage}\\m^2/m_0^2[F]:\begin{minipage}[t]{10cm}\raggedright\scriptsize ${+0.000}_{\times 6}$, ${+0.189}_{\times 8}$, ${+1.944}_{\times 4}$, ${+2.281}_{\times 3}$, ${+3.448}_{\times 4}$, ${+4.151}_{\times 3}$\end{minipage}\\m^2/m_0^2[\phi]:\begin{minipage}[t]{10cm}\raggedright\scriptsize ${-1.966}_{\times 4}$, ${-1.739}_{\times 9}$, ${-1.603}_{\times 9}$, ${-1.591}$, ${-1.180}_{\times 8}$, ${-0.981}_{\times 5}$, ${-0.731}_{\times 4}$, ${+0.000}_{\times 22}$, ${+0.582}$, ${+0.617}_{\times 5}$, ${+5.290}$, ${+6.721}$\end{minipage}\\
\end{longtable}}

{\small\begin{longtable}{L}{\bf S[{\scriptstyle 1/8}]01035834}: \mathfrak{so}(8)\rightarrow \mathfrak{u}(1),\quad \text{BF-stable}\\V/g^2\approx-10.3583429849,\quad |\nabla V/g^2|<{2.2}\cdot10^{-14}\\m^2/m_0^2[\psi]:\begin{minipage}[t]{10cm}\raggedright\scriptsize ${+1.062}$, ${+1.376}_{\times 2}$, ${+1.815}$, ${+1.825}$, ${+1.977}_{\times 2}$, ${+4.060}$\end{minipage}\\m^2/m_0^2[\chi]:\begin{minipage}[t]{10cm}\raggedright\scriptsize ${+0.011}$, ${+0.019}_{\times 2}$, ${+0.020}$, ${+0.030}_{\times 2}$, ${+0.036}_{\times 2}$, ${+0.233}_{\times 2}$, ${+0.262}_{\times 2}$, ${+0.299}$, ${+0.307}$, ${+0.348}_{\times 2}$, ${+0.371}$, ${+0.436}$, ${+0.560}$, ${+0.587}_{\times 2}$, ${+0.891}$, ${+0.932}_{\times 2}$, ${+0.972}_{\times 2}$, ${+1.201}$, ${+1.216}_{\times 2}$, ${+1.287}_{\times 2}$, ${+1.643}_{\times 2}$, ${+1.721}$, ${+1.822}$, ${+1.891}_{\times 2}$, ${+2.040}$, ${+2.476}_{\times 2}$, ${+3.014}$, ${+3.870}_{\times 2}$, ${+4.250}$, ${+4.790}$, ${+5.498}$, ${+5.504}_{\times 2}$, ${+5.892}_{\times 2}$, ${+6.180}$, ${+7.262}$, ${+7.299}$, ${+7.907}_{\times 2}$, ${+16.238}$\end{minipage}\\m^2/m_0^2[F]:\begin{minipage}[t]{10cm}\raggedright\scriptsize ${+0.000}$, ${+0.012}_{\times 2}$, ${+0.018}$, ${+0.067}_{\times 2}$, ${+0.073}_{\times 2}$, ${+0.157}_{\times 2}$, ${+0.417}$, ${+0.451}_{\times 2}$, ${+0.472}$, ${+1.709}$, ${+1.911}_{\times 2}$, ${+2.330}$, ${+2.713}_{\times 2}$, ${+2.879}$, ${+3.208}$, ${+3.320}$, ${+3.378}_{\times 2}$, ${+3.723}_{\times 2}$, ${+5.487}$\end{minipage}\\
\end{longtable}}

{\small\begin{longtable}{L}{\bf S[{\scriptstyle 1/8}]01400000}: \mathfrak{so}(8)\rightarrow \mathfrak{so}(3)+\mathfrak{so}(3),\quad \text{BF-stable}\\V/g^2\approx-14.0000000000,\quad |\nabla V/g^2|<{4.9}\cdot10^{-14}\\m^2/m_0^2[\psi]:\begin{minipage}[t]{10cm}\raggedright\scriptsize ${+2.143}_{\times 6}$, ${+3.857}_{\times 2}$\end{minipage}\\m^2/m_0^2[\chi]:\begin{minipage}[t]{10cm}\raggedright\scriptsize ${+0.000}_{\times 6}$, ${+0.655}_{\times 18}$, ${+4.488}_{\times 18}$, ${+8.571}_{\times 12}$, ${+15.429}_{\times 2}$\end{minipage}\\m^2/m_0^2[F]:\begin{minipage}[t]{10cm}\raggedright\scriptsize ${+0.000}_{\times 6}$, ${+1.714}_{\times 9}$, ${+4.286}_{\times 12}$, ${+8.571}$\end{minipage}\\m^2/m_0^2[\phi]:\begin{minipage}[t]{10cm}\raggedright\scriptsize ${-1.714}_{\times 20}$, ${-1.312}_{\times 9}$, ${+0.000}_{\times 22}$, ${+2.571}_{\times 9}$, ${+5.598}_{\times 9}$, ${+8.571}$\end{minipage}\\[6 pt]{\begin{minipage}[t]{10cm}\raggedright$\scriptstyle M_{00}{\approx}M_{11}{\approx}M_{22}{\approx}M_{33}{\approx}-M_{44}{\approx}-M_{55}{\approx}-M_{66}{\approx}-M_{77}{\approx}-M_{\dot0\dot4}{\approx}-M_{\dot1\dot6}{\approx}-M_{\dot2\dot7}{\approx}-M_{\dot3\dot5}\approx0.255201$\end{minipage}}\\[1ex]
\text{This~$\SO(3)\times \SO(3)$ equilibrium was first described in~\cite{Borghese:2013dja}.}
\end{longtable}}

\label{S:sol02270961}{\small\begin{longtable}{L}{\bf S[{\scriptstyle 1/8}]02270961}: \mathfrak{so}(8)\rightarrow \emptyset,\quad \text{BF-stable}\\V/g^2\approx-22.7096126583,\quad |\nabla V/g^2|<{8.4}\cdot10^{-12}\\m^2/m_0^2[\psi]:\begin{minipage}[t]{10cm}\raggedright\scriptsize ${+1.172}$, ${+3.366}_{\times 2}$, ${+3.551}$, ${+3.665}_{\times 2}$, ${+5.119}$, ${+5.632}$\end{minipage}\\m^2/m_0^2[\chi]:\begin{minipage}[t]{10cm}\raggedright\scriptsize ${+0.029}$, ${+0.035}_{\times 2}$, ${+0.073}$, ${+0.091}$, ${+0.111}$, ${+0.761}$, ${+0.923}$, ${+1.207}_{\times 2}$, ${+1.304}$, ${+1.384}_{\times 2}$, ${+2.419}$, ${+2.568}$, ${+3.143}_{\times 2}$, ${+3.615}$, ${+3.629}_{\times 2}$, ${+3.902}_{\times 2}$, ${+4.042}$, ${+4.377}$, ${+4.689}$, ${+5.677}$, ${+6.065}$, ${+6.103}_{\times 2}$, ${+7.021}_{\times 2}$, ${+7.071}$, ${+7.437}$, ${+7.650}_{\times 2}$, ${+7.667}$, ${+8.221}$, ${+8.572}$, ${+10.607}_{\times 2}$, ${+11.388}_{\times 2}$, ${+11.935}$, ${+12.018}_{\times 2}$, ${+12.196}$, ${+12.606}$, ${+12.648}$, ${+12.938}$, ${+13.463}_{\times 2}$, ${+14.205}$, ${+14.662}_{\times 2}$, ${+20.476}$, ${+22.529}$\end{minipage}\\m^2/m_0^2[F]:\begin{minipage}[t]{10cm}\raggedright\scriptsize ${+0.049}$, ${+0.274}_{\times 2}$, ${+1.037}$, ${+1.501}_{\times 2}$, ${+1.770}_{\times 2}$, ${+1.815}$, ${+1.884}$, ${+3.684}$, ${+3.732}$, ${+4.923}_{\times 2}$, ${+5.371}$, ${+5.521}_{\times 2}$, ${+5.631}_{\times 2}$, ${+6.192}$, ${+6.733}$, ${+6.813}$, ${+8.190}_{\times 2}$, ${+8.972}_{\times 2}$, ${+9.097}$, ${+9.251}$\end{minipage}\\
\end{longtable}}

\section{Trajectories in \texorpdfstring{$\omega$}{ω}-space}\label{sec:deformation_example}

\subsection{The \texorpdfstring{$\omega$}{ω}-trajectories of the scalar potential's equilibria}
\label{sec:trajectories}

It has been previously noted that vacuum solutions of the \SOom theories found for a specific value of $\omega$ do not simply disappear when we perturb the deformation parameter, but rather their position moves smoothly in field space and their physical properties (cosmological constant, mass spectra \textit{et cetera}) also change smoothly \cite{DallAgata:2012mfj,Borghese:2012qm,Borghese:2013dja}.
This suggests that we may choose a vacuum solution at some initial value of $\omega$ (such as $\omega=0$ or $\omega=\pi/8$) and follow it through the whole range $\omega\in[0,\pi/8]$.
This is not only a powerful technique to chart the landscape of vacuum solutions of \SOom gauged supergravity, but also reveals an intriguing web of relations between trajectories vacua.
It was already known that, following the $\omega$ trajectory of some vacua, they may in some cases move to the boundary of scalar field space, or in other cases exhibit the opposite behaviour and be entirely or partially independent of $\omega$.
In our numerical analysis, however, we find that $\omega$-trajectories can also intersect, branch, and stop at special values of the deformation parameter.

In the following sections we give a brief description of numerical approaches one can use to analyse the $\omega$-trajectories of vacua.
Then, we describe the results of applying such techniques to the recently discovered~$\SO(3)\;\mathcal{N}=1$ vacuum~S1384096 of the standard \SO(8) theory \cite{Bobev:2019dik} (within a $\bbZ_2^3$ invariant consistent truncation) and of the $\mathrm{\bf S1442018}$ vacuum \cite{Comsa:2019rcz} (in the full 70-dimensional scalar manifold).

\subsection{\texorpdfstring{$\omega$}{ω}-deformation via the Hessian of stationarity-violation}

Starting from either~$|\nabla V|^2$ or~$|Q_+|^2$ as a smooth measure of
stationarity-violation whose zeros correspond to the equilibria of the
supergravity vacuum equations of motion, cfr. equation \eqref{vac cond}, one idea to express
$\omega$-deformation of an equilibrium as an ordinary differential
equation starts by forming the~$71\times 71$ Hessian matrix of this
function.  The~$70\times 70$ sub-block that corresponds to parameter
variations with fixed~$\omega$ will have null eigenvalues due to an
equilibrium being physically equivalent to any other equilibrium on
its~$\SO(8)$-orbit. In addition to these ``goldstone'' directions (as
we will call them henceforth), there sometimes may be additional vanishing eigenvalues of the $70\times70$ Hessian.
This happens, for instance, for the critical points~S0800000 (with
residual~$\SU(4)^-$ symmetry) and~S1200000 (with residual~$\rmU(1)\times
\rmU(1)\;\mathcal{N}=1$ symmetry) of the standard \SO(8) theory as described in {\cite{Comsa:2019rcz}}.
Now, if a given equilibrium admits~$\omega$-deformation, this will show in the form of the~$71\times 71$-Hessian having an additional null eigenvector beyond those that do not involve changing~$\omega$. 
So, by isolating this extra null eigenvector (if it exists), unit-normalizing it, and resolving the direction ambiguity by fixing the sign of its~$\omega$-coordinate, we obtain an ordinary differential equation~(ODE) that schematically is of the form
\begin{equation}\label{ODE}
\frac{\dd}{\dd s}(\,\phi_0(s),\ldots,\phi_{69}(s);\omega(s)\,)=
\vec{f}(\phi_0(s),\ldots,\phi_{69}(s);\omega(s))\,,
\end{equation}
where $\vec{f}$ is the null 71-dimensional eigenvector of the~$71\times 71$-Hessian we just discussed and $s$ parameterises the curve in $\omega$ and field space.
While this approach can indeed be made to work, and TensorFlow is
powerful enough to synthesize a computational graph that requires a
third derivative (fourth derivative if we also want the ODE's
Jacobian) of a somewhat large complex matrix
exponential\footnote{Two derivatives for the Hessian, and a third derivative if
stationarity-violation was defined in terms of~$|\nabla V|^2$}, this
naive strategy has major disadvantages, both conceptually and
computationally. The biggest problem here is that, as in particular
the example discussed in section~\ref{sec:deformation_example} will
show, it can easily happen that, as we move along a
$\omega$-trajectory, the~$\omega$-coordinate of the 71-dimensional
unit tangent vector shrinks to zero. While this has not
been described beforehand in the literature, it turns out to be
a rather common phenomenon that, as we follow an~$\omega$-deformation
trajectory in~$71$-dimensional parameter space, we reach a point
where~$\omega$ attains an extremum, and any attempt to continue
further along the trajectory requires changing the sign
of~$d\omega/ds$ at such a point with~$d\omega/ds=0$.
So, the overall scheme sketched in this section would require
further amendments in order to also handle such situations.
As a superior computational strategy exists, which will
be discussed next, we do not explain these amendments in detail.

\subsection{\texorpdfstring{$\omega$}{ω}-deformation via the Jacobian of the potential}

Observing that additional flat, non-Goldstone directions of the scalar potential are less of a computational obstacle than~$\omega$-extrema, and also realizing that relevant properties of the Hessian of the stationarity-violation have more directly accessible counterparts in the Jacobian of the potential's gradient in the local frame, a more powerful approach to the numerical study of $\omega$-deformation works as follows.
Starting from the $T$-tensor \eqref{ttensor} as a function of~71 parameters (the scalar field vevs and $\omega$), we compute the fermion shifts and hence the self-dual tensor~$Q_+^{ijkl}$ that has to vanish at a critical point.
We can arrange its components in terms of a 70-dimensional real vector denoted by $(Q_+)_n$, $n=0,\ldots,69$.
We then take its derivative with respect to the scalar field vevs and also with respect to $\omega$, thus constructing a $70\times71$ Jacobian matrix as follows
\begin{equation}\label{J of Q}
J_{np}:=\partial_p\,(Q_+)_n,\quad 
\partial_p = \frac\partial{\partial\phi^p}\,,\text{ if } p<70\,,\quad 
\partial_{70} = \frac\partial{\partial\omega}\,,
\end{equation}
where we are denoting the scalar field vevs by $\phi^0,\ldots,\phi^{69}$ for simplicity.

Let us now suppose that we have found a vacuum solution for some initial value $\omega_*$ of the deformation parameter and with scalar field vevs $\phi_*^0,\ldots,\phi_*^{69}$.
The Goldstone modes correspond to the \SO(8)-orbit of these vevs, which by gauge invariance preserves the vacuum condition $(Q_+)_n=0$.
We identify the infinitesimal Goldstone modes by applying the 28~generators of $\SO(8)$ to the vector $\phi_*^n$, $n=0,\ldots,69$, constructing a $28\times70$ dimensional matrix of rank equal to the number of broken gauge symmetries,
\begin{equation}
G_{\cA}{}^n = (T_{\cA})^n{}_m \phi_*^m
\end{equation}
where $(T_{\cA})^n{}_m$ are the \SO(8) generators in the $\mathbf{35}_s+\mathbf{35}_c$ and the $\cA$ is an adjoint \so(8) index.
We may trivially extend the index $n$ to $n=70$ by adding a vanishing column to the matrix.
We then obviously have $G_{\cA}{}^p J_{np} = 0 $.
Projecting the 71-dimensional index of $J_{np}$ on the (right) kernel of $G_{\cA}{}^p$, we effectively eliminate the Goldstone directions.\footnote{In order to have better numerical accuracy, one first identifies a basis for the Goldstone directions from a singular value decomposition of the matrix $G_\cA{}^n$ and then performs the projection on the Jacobian.}
Let us denote by $\tilde J_{np}$ such projected Jacobian, where now $p=0,\ldots,(28-\dim\rmG_{\rm res})$.
The right kernel\footnote{
This again can be computed for via a singular value decomposition, selecting the vanishing singular values. Computationally, given that numerical calculations here are intrinsically noisy, and the magnitude of the relevant singular values of~$\tilde J_{nk}$ often spans multiple orders of magnitude, it makes sense to introduce a prescription that considers a singular value to be effectively zero if it is orders of magnitude below the geometric mean of the top third of largest singular values or so.} 
of $\tilde J_{np}$  identifies infinitesimal flat directions in the vacuum condition that do not correspond to Goldstone modes.
Elements of this kernel whose $\omega$ component vanishes correspond to massless scalar modes at the vacuum.
We are interested in elements with a non-zero $\omega$ component.
Such elements indicate the possbility of infinitesimally changing the value of $\omega$ while at the same time correcting the scalar field vevs so that the vacuum condition remains satisfied. The value of the cosmological constant need not stay constant along this deformation.
If this kernel has dimension higher than 1 and is not orthogonal to the $\omega$ direction, we might be able to continue an~$\omega$-trajectory along multiple different paths: unless higher order corrections prevent us from maintaining the stationarity condition~$(Q_+)_n=0$, we are at a crossing of $\omega$-trajectories.
Examples for such crossings are described in the next subsections.

\subsection{ODE integration}

The procedure sketched in the previous section allows us to efficiently generate an infinitesimal $\omega$-deformation of a given vacuum.
To find a finite deformation and follow a vacuum in $\omega$ space, we still need to numerically integrate an ODE such as \eqref{ODE}, but now we can use a (unit normalised) element of the right kernel of $\tilde J$ with non-vanishing $\omega$ component in place of the vector $\vec f$ computed from the $71\times71$ Hessian in the previous section.
Some extra care must be taken because we the want to maintain the vacuum condition $(Q_+)_n=0$ along the entire trajectory, despite numerical rounding errors gradually accumulating as we perform ODE-integration.

Furthermore, at points where we have with multiple options to continue an~$\omega$-trajectory, we would like to have a well-defined ``default prescription'' to go through the crossing. 
The most straightforward choice is to use the parameter update-step from the previous coordinate-update done by ODE integration, project it (using orthonormality) onto the subspace of infinitesimal coordinate-changes which preserve $(Q_+)_n=0$, and afterwards normalize to unit length. This also addresses the problem that unit-normalized singular vectors (i.e., elements of the right kernel of $\tilde J_{np}$, computed via singular value decomposition) may sign-flip between one ODE integration step and the next.

Conceptually, it hence seems to make sense to rather
think of the ODE as a differential algebraic equation (DAE). 
The prescription to handle crossings that was described in the previous
section asks for keeping track of the current marching direction, and
as implementing this requires either major modifications to the ODE,
or small adjustments in the ODE/DAE solver, we here use a rather
simplistic procedure, a modified Runge-Kutta RK4 ODE integration that
also provides information about the previous marching direction as a
tie-breaker at crossings. This should be seen as a first step towards
further innovation which then should also tackle automatic step size
adjustment to in particular deal with closely spaced crossings.
Whenever the numerical violation of the stationarity condition becomes
too large (such as: numerically~$>10^{-12}$), we interrupt
ODE-integration and insert a minimization step that brings this
violation back down to numerically-zero.  Given that we still want to
retain information about the previous marching direction when
continuing ODE-integration after such a `numerical quality
improvement' step, care has to be taken with this optimization step to
ensure that we do not accidentally drift away too far on the
$\SO(8)$-orbit of the solution.

\subsection{Example omega trajectories}

\begin{figure}[hbt!]
\centering
\includegraphics[width=0.75\textwidth]{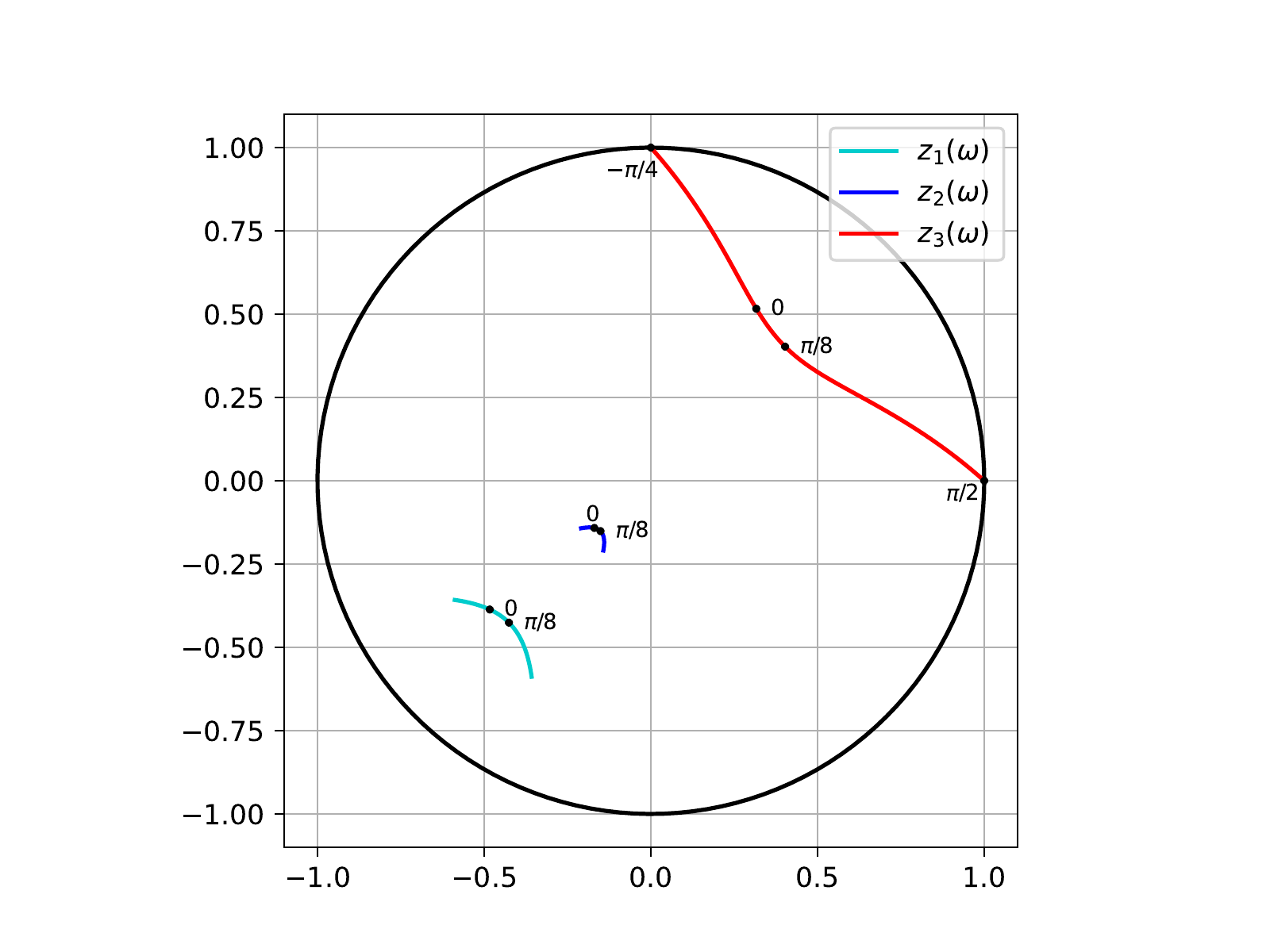}\\
\includegraphics[width=0.4\textwidth]{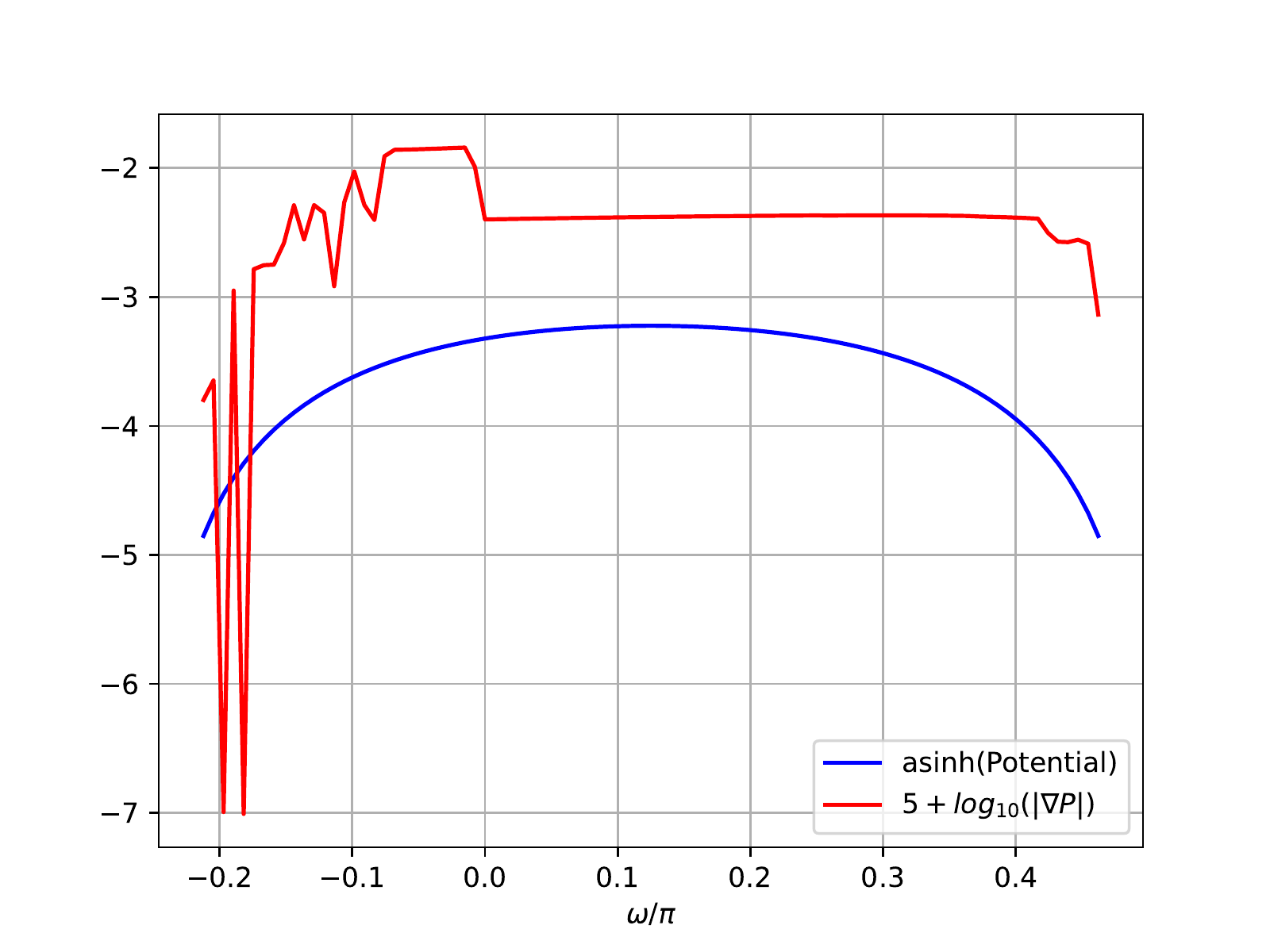}\kern1cm
\includegraphics[width=0.4\textwidth]{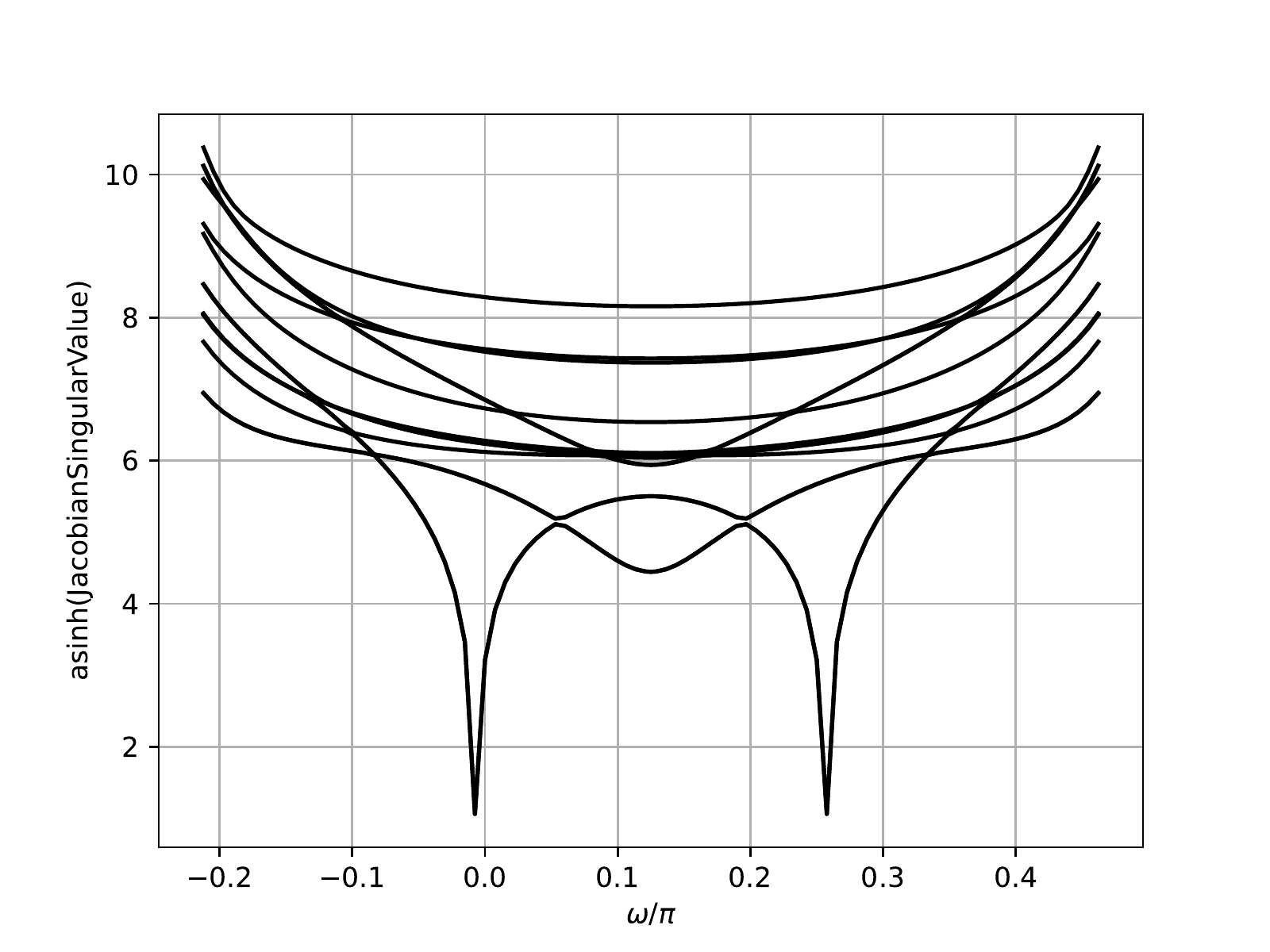}\\
\caption{$\omega$-deformation of the $\mathrm{SO}(3)\;\mathcal{N}=1$ equilibrium of the de Wit-Nicolai model on the $\mathcal{M}_{14}:=(\mathrm{SU}(1,1)/\mathrm U(1))^{7}$ submanifold of~$\mathcal{M}_{70}$. Top: Evolution of the~$z_{1,2,3}$ parameters on the Poincare disc as a function of~$\omega$. The $\omega$-trajectory is symmetric around~$\omega=\pi/8$. Left: asinh-``squashed'' Potential and shifted log-scale stationarity-violation (gradient of the potential). This diagram shows that the potential is monotonic in~$|\omega|$, and that ODE-integration always kept the numerical stationarity-violation $<10^{-6}$, indicating reasonably good numerical quality of results. Right: asinh-squashed singular values of the $70\times14$ Jacobian of the potential. Both the level-crossing avoidance and the observation that singular values do not reach down to zero are limited-resolution effects.} 
\label{fig:so3n1_flow}
\end{figure}
The procedure sketched above is readily generalized to cases where we want to constrain the trajectory to some lower-dimensional submanifold of $\cM_{70}$, although caution has to be exercised with then potentially becoming blind to situations where the trajectory may want to leave that submanifold at some point. %
Then, as a first example of $\omega$-trajectories of vacuum solutions, we study the~$\omega$-deformation of the recently discovered~$\SO(3)\;\mathcal{N}=1$ vacuum~S1384096 of the standard \SO(8) theory \cite{Bobev:2019dik} on the 14-dimensional submanifold~$\cM_{14}=\left(\SU(1,1)/\rmU(1)\right)^7$.\footnote{
The truncation to this submanifold has been discussed in many places in the literature. See for instance \cite{Aldazabal:2006up,DallAgata:2005zlf,Catino:2013ppa} and references therein.}
The truncation to $\cM_{14}$ has the advantage that the trajectory can be easily depicted and we do so in figure~\ref{fig:so3n1_flow}.
This scalar manifold can be parametrised in terms of seven complex coordinates $\zeta_k$ restricted to the unit circle, such that the supergravity K\"ahler potential reads $\cK = - \sum_{k=1}^7\log(1-\zeta_k\bar\zeta_k)$. 
This coordinate parametrization is described in equations~$(7.1)-(7.7)$ in~\cite{Bobev:2019dik}.
In particular, we shall restrict to an \SO(3) invariant subsector given by
\begin{equation}
\zeta_1 = -z_2\,,\qquad
\zeta_2 = \zeta_6=\zeta_7 = -z_3\,,\qquad
\zeta_3=\zeta_4=\zeta_5=z_1\,.
\end{equation}

An important feature of this deformation-trajectory, which here is symmetric around~$\omega=\pi/8$, is that $z_3$ approaches the boundary of the unit disk in the limits~$\omega\to-\pi/4$ and~$\omega\to\pi/2$, where we also observe~$V/g^2\to-\infty$.
Looking at the plot of singular values of the Jacobian we also observe that there are two special values of $\omega$, symmetric around~$\omega=\pi/8$ and corresponding to the two dips in the picture, where the potential acquires additional flat directions on~$\mathcal{M}_{14}$. 
Along the depicted trajectory we encounter equilibria that are equivalent to the vacua ${\bf S[{\scriptstyle 1/8}]01253393}$, ${\bf S1384096}$ and ${\bf S[{\scriptstyle 1/8}]02150686}$. 
At $\omega/\pi\approx1/8\pm0.132$, we find an equilibrium with $\mathfrak{so}(3)$ symmetry, $V/g^2\approx-14.01$, and in total 30 `mass-zero' scalars.
Furthermore, calculating the same trajectory on the entire scalar manifold~$\mathcal{M}_{70}$ shows that there is another equilibrium with extra flat directions at $\omega/\pi\approx1/8\pm0.246$, $\mathfrak{so}(3)$ symmetry, $V/g^2\approx-20.97$, and also in total 30~`mass-zero' scalars. 
In this latter case, none of the additional flat directions intersect~$\mathcal{M}_{14}$.

As a second example of $\omega$ trajectories, we now take a look at the $\omega$-deformation trajectory of a specific equilibrium that was picked to exemplify much of the generic behavior that can occur under such deformation. This shows multiple new features that have not been encountered in earlier such studies of more accessible critical points with a high degree of residual symmetry, such as~\cite{Borghese:2012qm,DallAgata:2012plb,Borghese:2012zs,Borghese:2013dja,Catino:2013ppa,Gallerati:2014xra}.
This computation has been performed numerically on the full $70$-dimensional scalar manifold, via ODE integration as described in the previous sections. 
While it is in principle possible that step sizes might have been taken too large, and ``jumping over a point of special interest'' or ``jumping onto a different trajectory'' may have happened, the observed length scales of relevant features make this implausible here.
Nevertheless, as we will see, it is especially the start of the $\omega$-trajectory described below that serves as a warning that such investigations have to be done with great diligence.
Along the entire trajectory, the length of the gradient of the potential (taken with respect to the local coordinate frame on the scalar manifold) stayed numerically small ($<10^{-10}$).

\begin{figure}[hbt!]
\centering
\includegraphics[width=0.95\textwidth]{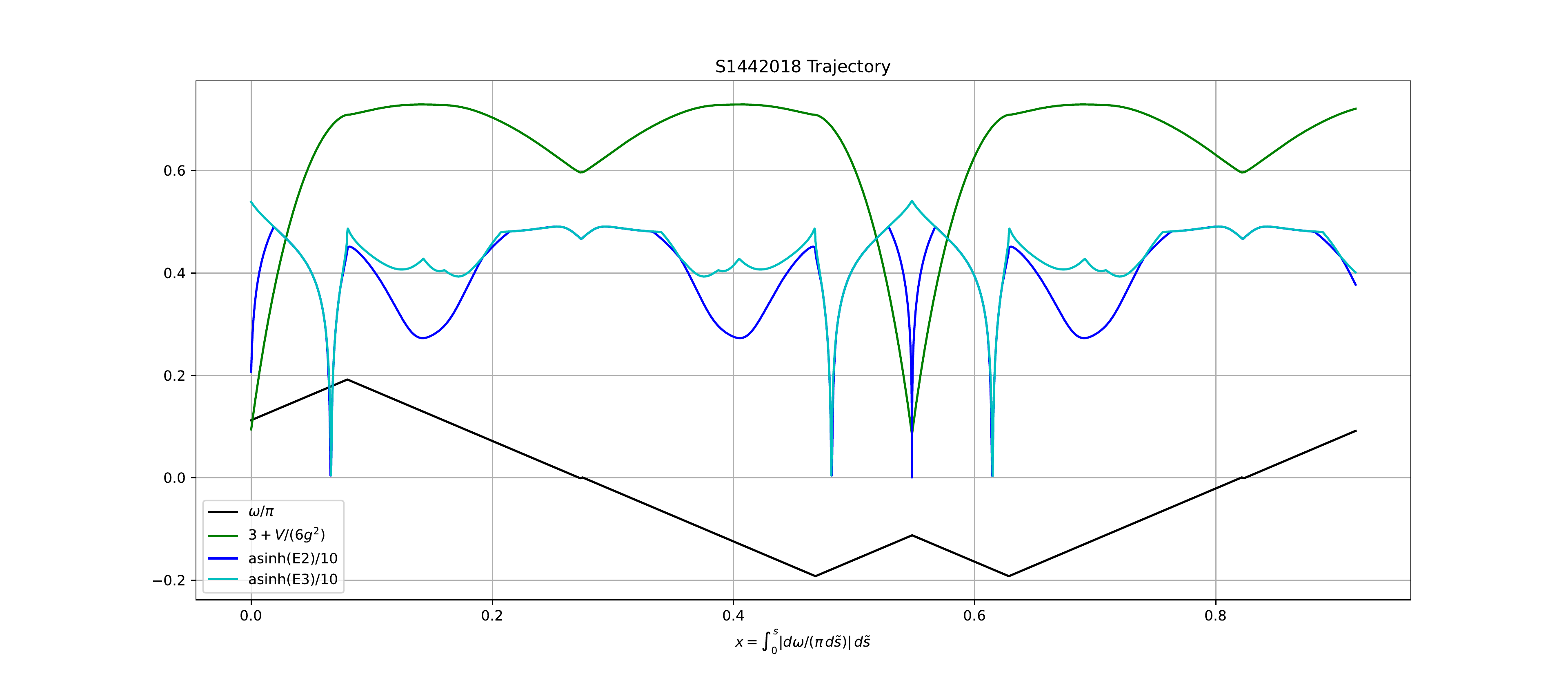}\\
\includegraphics[width=0.45\textwidth]{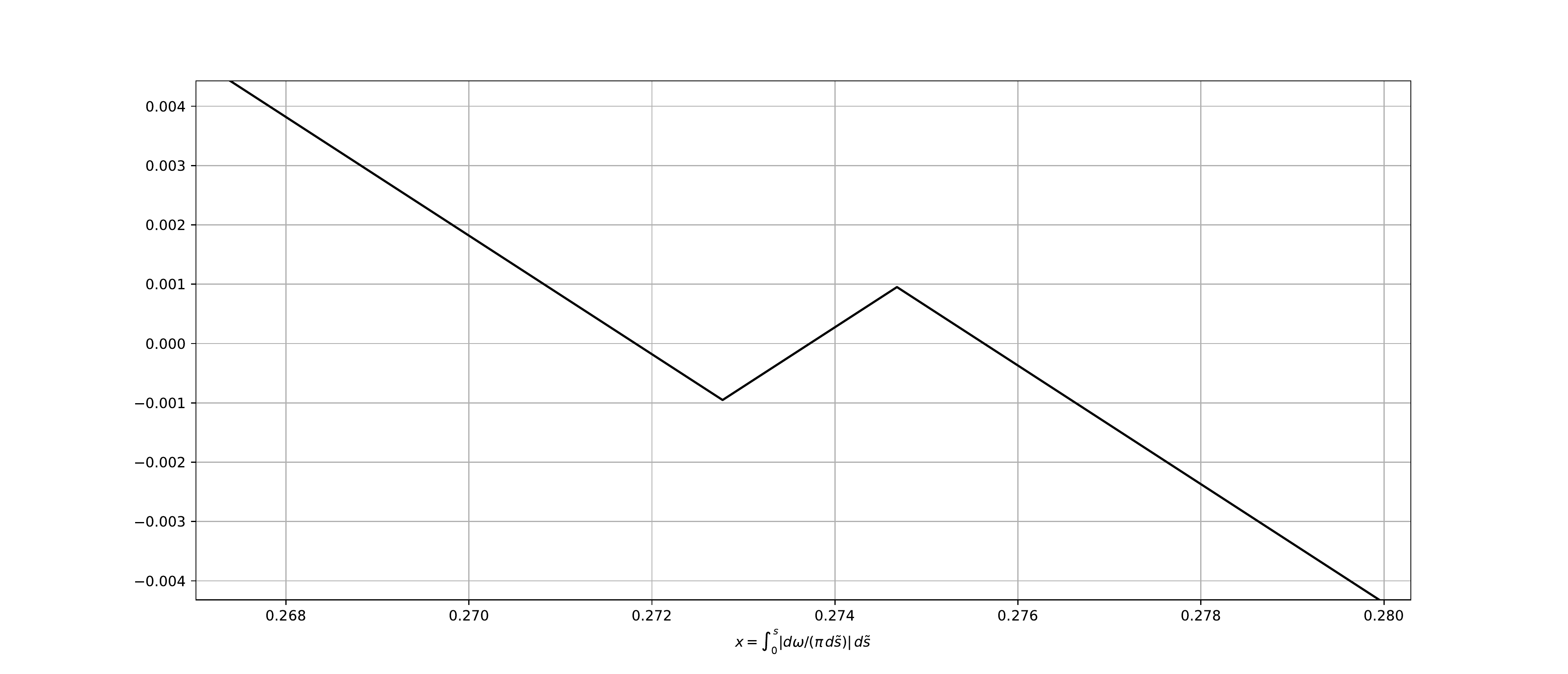}\kern1em%
\includegraphics[width=0.45\textwidth]{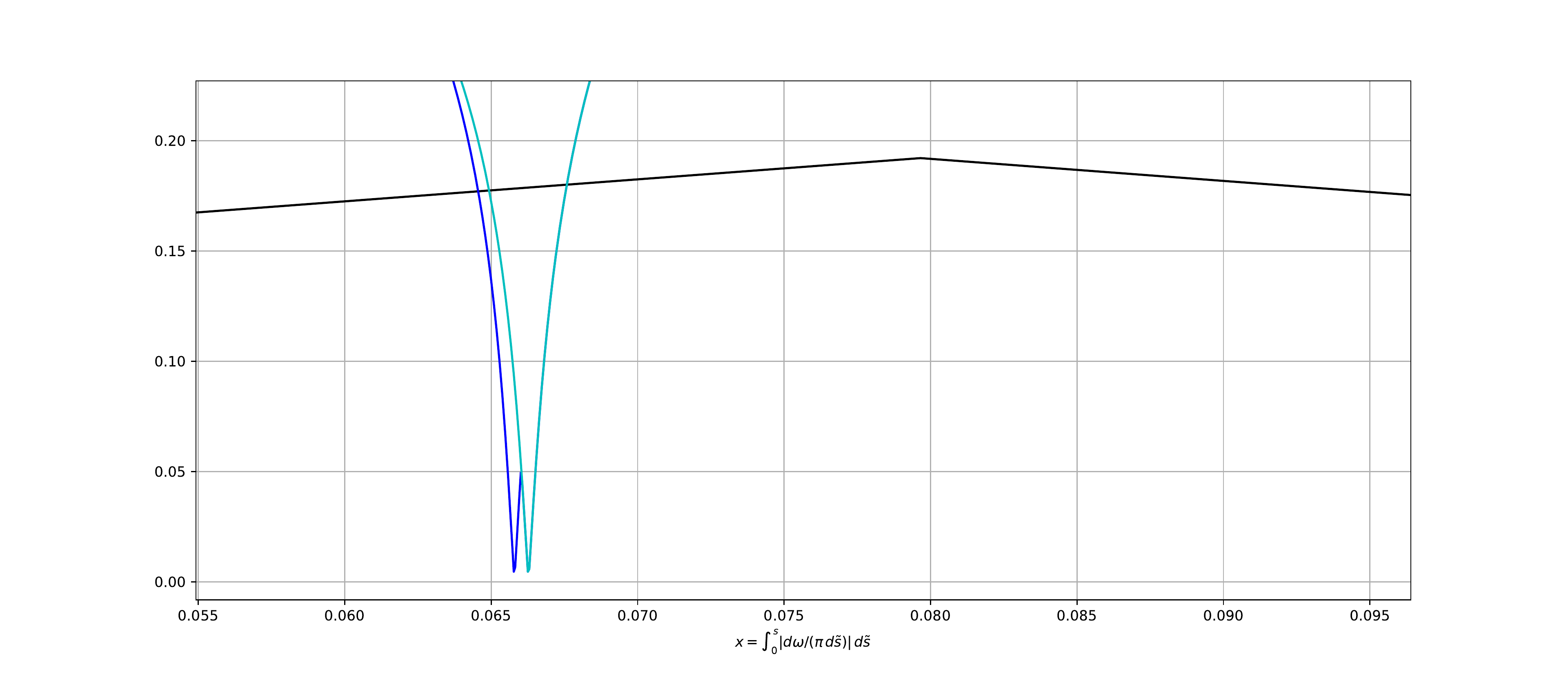}\\
\caption{$\omega$-deformation of the S1442018 equilibrium of the de Wit-Nicolai model.}
\label{fig:S1442108_trajectory}
\end{figure}

\begin{figure}[hbt!]
\centering
\includegraphics[width=0.75\textwidth]{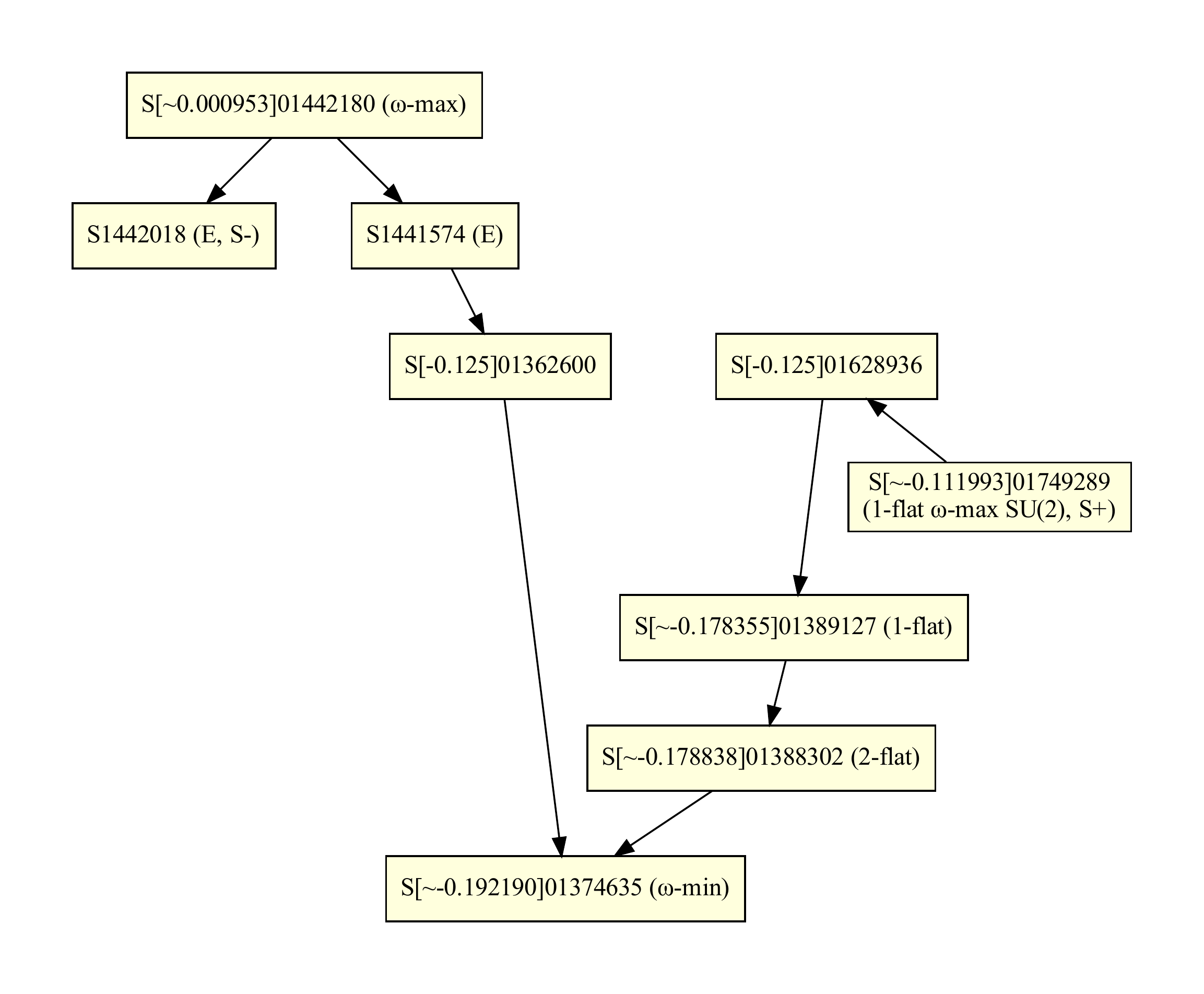}\\
\small\begin{longtable}{|L|R|L|C|C|C|C|}
\hline
\text{Name}&\text{Potential}&|\nabla V/g^2|&\text{Symmetry}&\text{Type}\\
\hline
{\bf S1442018}&-14.4201873779&<{1.7}\cdot10^{-13}&\mathfrak{u}(1)&\omega=0\\
{\bf \somega{\approx0.000953}{01442180}}&-14.4218086363&<{1.8}\cdot10^{-13}&\mathfrak{u}(1)&\omega-\text{Maximum}\\
{\bf S1441574}&-14.4157405138&<{2.7}\cdot10^{-13}&\mathfrak{u}(1)&\omega=0\\
\somega{-1/8}{01362600}&-13.6260039293&<{1.3}\cdot10^{-12}&\mathfrak{u}(1)&\omega=-\pi/8\\
\somega{\approx-0.192190}{01374635}&-13.7463556328&<{1.4}\cdot10^{-12}&\mathfrak{u}(1)&\omega-\text{Minumum}\\
\somega{\approx-0.178838}{01388302}&-13.8830231447&<{7.5}\cdot10^{-12}&\mathfrak{u}(1)&+2\;\text{flat}\\
\somega{\approx-0.178355}{01389127}&-13.8912772896&<{5.6}\cdot10^{-13}&\mathfrak{u}(1)&+1\;\text{flat}\\
\somega{-1/8}{01628936}&-16.2893639331&<{1.9}\cdot10^{-13}&\mathfrak{u}(1)&\omega=-\pi/8\\
\somega{\approx-0.111993}{01749289}&-17.4928962398&<{2.2}\cdot10^{-13}&\mathfrak{so}(3)&\omega-\text{Maximum}, +1\;\text{flat}\\
\hline
\end{longtable}
\caption{Schematic structure of the S1442018-equilibrium's $\omega$-deformation trajectory. Arrows point in the direction of \emph{decreasing}~$\omega$.  `E'=`Non-Dyonic gauging', `S-'=`Trajectory continues symmetrically under a sign change of $d\omega/ds$', `S+'=`Trajectory continues symmetrically around this $\omega$-extremum'.}
\label{fig:S1442108_graph}
\end{figure}

Figure~\ref{fig:S1442108_trajectory} shows the scalar potential
(re-scaled and shifted to make it align better with the other
displayed data), the $\omega$-parameter (which by construction
here always has slope $\pm 1$, see below), as well as the two
lowest-lying singular values\footnote{As a function of the
curve-parameter, the generalized eigenvectors of the Jacobian
are of course described by smooth functions, and the associated
singular values are likewise smooth, but due to crossing with
not-shown eigenvalues, we observe `kinks' in the graphs of these
singular values of the Jacobian.}
from the singular value decomposition of the Jacobian from eq.~\ref{J of Q}.
The first curious new phenomenon here we encounter in the form of the
need to describe the trajectory in terms of `distance walked along
$\omega$-direction', i.e. $x=\int_0^s|d\omega/d\tilde s|\,d\tilde s$,
as adjusting $\omega$ will lead us to points where $\omega$ cannot be
increased (respectively, decreased) any further while maintaining a
length-zero gradient of the scalar potential (along the 70 scalar
coordinates). Closer inspection shows that at such points, the
trajectory typically `bends around' and can be continued (by following
its tangent vector in 71-parameter space) with a sign change in
$d\omega/ds$. 
Figure~\ref{fig:S1442108_graph} shows a `subway map' of the relevant
part of the trajectory, which has to be mirrored once at its final
endpoint, and then mirrored in its new entirety once again, replacing
$\omega\to\-\omega$, to form the full loop.
The role of the special points on this trajectory is described
in the following.

Apart from the $\omega$-extrema on the trajectory that also can be
viewed as two $\omega$-deformation trajectories connecting smoothly,
there also are points where the potential develops relevant extra flat
directions (beyond those expected from symmetry breaking). These show up
in the form of one or multiple singular values of the relevant part of
the Jacobian dipping down to zero (not perfectly so in the plot, due
to sampling effects). Depending on whether or not flatness also
persists to higher order, these special points with extra flat
directions may or may not be places where multiple \emph{different}
$\omega$-deformation trajectories meet. The prescription used to
integrate through such higher order critical points of the scalar
potential is to ODE-integrate the unit tangent vector obtained by
(71-parameter-)orthogonal projection of the previous step's unit
tangent vector onto the subspace of flat directions and subsequent
rescaling to bring the length back to~1. As long as
$\omega$-deformation admits only one direction along which the
stationarity condition can be maintained, this prescription will
merely trace out the trajectory, but at higher critical points, it
will `take the straight path through it'.  This table shows the
sequence of relevant points that we encounter along the trajectory as
determined with this prescription for handling points with additional
flat directions.

\newpage

Starting at a~$\omega=0$ point on the shown trajectory, we encounter
equivalent situations along both directions, the physics is symmetric
with respect to exchanging $\omega\to-\omega$. However, when going in
the direction of increasing $\omega$, we already at
$\omega/\pi\approx0.000953$ encounter a point where $\omega$ cannot be
increased further and we have to turn around. Continuing further along
this trajectory, we pass through $\omega=0$ again, where we find
another already known equilibrium of the de Wit-Nicolai model,
S1441574. Pushing further, we first meet an instance of
$\somega{1/8}{01362600}$ at $\omega/\pi=-1/8$ and then encounter another point
where we have to turn around at $\omega/\pi\approx-0.1922$.  The
maximum of the scalar potential is not at $\omega=-\pi/8$, but near
$\omega/\pi\approx-0.13$, with $V/g^2\approx-13.6254$ (this point is
not shown in detail). Soon after, we appear to run into a pair of
closely spaced higher-degree critical points, the first one at
$\omega/\pi\approx-0.1788$ having two, and the second one at
$\omega/\pi\approx-0.1784$ having one additional flat direction. It is
conceivable, but unlikely, that this may actually be a single critical
point of higher degree, which due to accuracy limitations of the
numerical method here shows as two closely spaced such points --
numerics suggests that the properties of these two distinct points can
be obtained to high accuracy and are not overly affected by
noise. (The rather different near-zero scalar masses also seem to
confirm this picture.) Afterwards, with $\omega/\pi$ still increasing,
we cross $\omega/\pi=-1/8$ again at a negative-$\omega$ equivalent of
the equilibrium~$\somega{1/8}{01628936}$, and finally encounter another point
with an extra flat direction that additionally does not allow us to
increase~$\omega$ any further at $\omega/\pi\approx-0.112$.

Remarkably, `extra flat direction' at this point is with respect to an
enlarged gauge symmetry, which no longer is~$\rmU(1)$, as along the rest
of the trajectory, but rather~$\SU(2)$. Further inspection -- which
goes beyond the scope of the present article -- shows that each of
these points with additional flat directions is a crossing
of~$\omega$-trajectories. The $\SU(2)$-symmetric point
$\somega{\approx-0.111993}{01749289}$ admits $\omega$-deformation along
a trajectory with residual symmetry $\SU(2)$, which encounters S1271622
at~$\omega/\pi=-1/4$ and is symmetric around this point.
The potential on this trajectory diverges towards~$-\infty$
in the~$\omega\to0$ limit. The point $\somega{\approx-0.178838}{01388302}$
with two additional flat directions also lies on a~$\omega$-trajectory
that is symmetric around this point, where it attains its maximal
value of $\omega$. Along this trajectory, which has no residual
Lie symmetry apart from this point,~$\omega/\pi$ oscillates in the
interval~$[-0.322\ldots-0.178]$. At $\omega/\pi=-1/4$, we encounter
the known equilibrium~S1367611, and the physics is also symmetric
around this point under exchange $\omega\to-\omega$.
Very close to the~$\omega$-extremum at~$\omega/\pi\approx-0.17883790$,
we apparently encounter another point with one additional flat direction
at $\omega/\pi\approx-0.17899393$ that also might be a crossing, but again,
this feature may actually be related to ODE-integration not navigating the
crossing in an entirely straight way.

Numerical investigations also show that the point
$\somega{\approx-0.178355}{01389127}$, at which we find one additional flat
direction, also lies on a trajectory along which~$\omega/\pi$
oscillates in the interval~$[-0.214\ldots-0.178]$. The corresponding
$\omega$-minumum at~$\omega/\pi\approx0.214$ also has one additional
flat direction and hence might be another crossing point.

Details about the properties of the special points on the
S1442018-trajectory (but not the trajectories reached by taking a turn
at a crossing) {are shown below}, unless their properties already are
described elsewhere in this article (such as for
$\somega{1/8}{01362600}$), or in earlier work (such as for
S1441574). For every type of special point, we list only one
representative of the physics.


\label{S:sol[0.000953]1442180}{\small\begin{longtable}{L}\somega{\approx0.000953}{01442180}: \mathfrak{so}(8)\rightarrow \mathfrak{u}(1),\quad \text{BF-unstable}\\V/g^2\approx-14.4218086363,\quad |\nabla V/g^2|<{1.8}\cdot10^{-13}\\m^2/m_0^2[\psi]:\begin{minipage}[t]{10cm}\begin{flushleft}\scriptsize ${+2.438}_{\times 2}$, ${+2.757}_{\times 2}$, ${+3.764}_{\times 2}$, ${+3.910}_{\times 2}$\end{flushleft}\end{minipage}\\m^2/m_0^2[\chi]:\begin{minipage}[t]{10cm}\begin{flushleft}\scriptsize ${+0.010}_{\times 2}$, ${+0.022}_{\times 2}$, ${+0.208}_{\times 4}$, ${+0.356}_{\times 2}$, ${+0.375}_{\times 2}$, ${+0.402}_{\times 2}$, ${+0.416}_{\times 2}$, ${+1.062}_{\times 2}$, ${+1.267}_{\times 2}$, ${+2.636}_{\times 2}$, ${+2.676}_{\times 2}$, ${+4.293}_{\times 2}$, ${+5.442}_{\times 2}$, ${+5.948}_{\times 2}$, ${+6.040}_{\times 2}$, ${+7.056}_{\times 2}$, ${+7.213}_{\times 2}$, ${+7.943}_{\times 2}$, ${+8.411}_{\times 2}$, ${+9.751}_{\times 2}$, ${+9.791}_{\times 2}$, ${+10.062}_{\times 2}$, ${+11.028}_{\times 2}$, ${+12.048}_{\times 2}$, ${+14.276}_{\times 2}$, ${+15.055}_{\times 2}$, ${+15.641}_{\times 2}$\end{flushleft}\end{minipage}\\m^2/m_0^2[F]:\begin{minipage}[t]{10cm}\begin{flushleft}\scriptsize ${+0.000}$, ${+0.049}$, ${+0.099}_{\times 2}$, ${+0.235}$, ${+0.465}_{\times 2}$, ${+0.492}$, ${+2.515}_{\times 2}$, ${+2.713}$, ${+3.874}$, ${+3.880}_{\times 2}$, ${+4.101}_{\times 2}$, ${+4.241}$, ${+4.417}$, ${+4.684}$, ${+4.762}$, ${+5.240}$, ${+5.562}$, ${+6.456}$, ${+7.319}_{\times 2}$, ${+7.732}$, ${+9.273}$, ${+9.938}$\end{flushleft}\end{minipage}\\m^2/m_0^2[\phi]:\begin{minipage}[t]{10cm}\begin{flushleft}\scriptsize ${-3.723}$, ${-3.443}$, ${-3.067}$, ${-2.406}$, ${-2.171}_{\times 2}$, ${-2.001}$, ${-1.916}$, ${-1.481}_{\times 2}$, ${-1.358}_{\times 2}$, ${-1.265}_{\times 2}$, ${-1.162}$, ${-0.991}_{\times 2}$, ${-0.822}$, ${-0.482}$, ${-0.433}_{\times 2}$, ${+0.000}_{\times 28}$, ${+1.440}_{\times 2}$, ${+1.779}$, ${+2.710}$, ${+3.013}_{\times 2}$, ${+3.822}_{\times 2}$, ${+5.496}$, ${+6.544}$, ${+6.886}$, ${+7.461}$, ${+8.470}_{\times 2}$, ${+8.803}$, ${+10.161}$, ${+10.471}$, ${+11.450}_{\times 2}$, ${+12.729}$, ${+13.316}$\end{flushleft}\end{minipage}\\[6 pt]{\scriptstyle\begin{minipage}[t]{10cm}\begin{flushleft}$M_{00}{\approx}M_{11}\approx0.167798$, $M_{22}\approx-0.234306$, $M_{23}\approx0.000551$, $M_{26}\approx-0.000826$, $M_{27}\approx0.017379$, $M_{33}\approx-0.028997$, $M_{36}\approx-0.307808$, $M_{37}\approx-0.000065$, $M_{44}\approx0.225725$, $M_{55}\approx-0.085271$, $M_{66}\approx0.048887$, $M_{67}\approx0.000018$, $M_{77}\approx-0.261634$, $M_{\dot0\dot0}{\approx}M_{\dot1\dot1}\approx0.436281$, $M_{\dot2\dot2}{\approx}M_{\dot3\dot3}\approx0.056447$, $M_{\dot4\dot4}{\approx}M_{\dot5\dot5}\approx-0.088405$, $M_{\dot6\dot6}{\approx}M_{\dot7\dot7}\approx-0.404322$\end{flushleft}\end{minipage}}\\\end{longtable}}


\label{S:sol[-0.192190]1374635}{\small\begin{longtable}{L}\somega{\approx-0.192190}{01374635}: \mathfrak{so}(8)\rightarrow \mathfrak{u}(1),\quad \text{BF-unstable}\\V/g^2\approx-13.7463556328,\quad |\nabla V/g^2|<{1.4}\cdot10^{-12}\\m^2/m_0^2[\psi]:\begin{minipage}[t]{10cm}\begin{flushleft}\scriptsize ${+2.037}_{\times 2}$, ${+2.889}_{\times 2}$, ${+3.657}_{\times 2}$, ${+4.151}_{\times 2}$\end{flushleft}\end{minipage}\\m^2/m_0^2[\chi]:\begin{minipage}[t]{10cm}\begin{flushleft}\scriptsize ${+0.017}_{\times 2}$, ${+0.018}_{\times 2}$, ${+0.110}_{\times 2}$, ${+0.163}_{\times 2}$, ${+0.168}_{\times 2}$, ${+0.281}_{\times 2}$, ${+0.387}_{\times 2}$, ${+0.397}_{\times 2}$, ${+0.849}_{\times 2}$, ${+1.422}_{\times 2}$, ${+1.563}_{\times 2}$, ${+2.835}_{\times 2}$, ${+3.841}_{\times 2}$, ${+3.905}_{\times 2}$, ${+4.681}_{\times 2}$, ${+5.946}_{\times 2}$, ${+6.670}_{\times 2}$, ${+7.097}_{\times 2}$, ${+8.053}_{\times 2}$, ${+8.146}_{\times 2}$, ${+8.782}_{\times 2}$, ${+9.973}_{\times 2}$, ${+10.516}_{\times 2}$, ${+11.084}_{\times 2}$, ${+11.556}_{\times 2}$, ${+14.626}_{\times 2}$, ${+16.603}_{\times 2}$, ${+17.503}_{\times 2}$\end{flushleft}\end{minipage}\\m^2/m_0^2[F]:\begin{minipage}[t]{10cm}\begin{flushleft}\scriptsize ${+0.000}$, ${+0.072}_{\times 2}$, ${+0.087}$, ${+0.149}$, ${+0.310}_{\times 2}$, ${+0.320}$, ${+2.136}$, ${+2.545}$, ${+3.042}_{\times 2}$, ${+3.092}_{\times 2}$, ${+3.155}$, ${+3.627}_{\times 2}$, ${+4.243}$, ${+4.347}$, ${+4.992}$, ${+5.060}$, ${+5.147}$, ${+7.164}_{\times 2}$, ${+8.891}$, ${+9.253}$, ${+9.823}$, ${+10.030}$\end{flushleft}\end{minipage}\\m^2/m_0^2[\phi]:\begin{minipage}[t]{10cm}\begin{flushleft}\scriptsize ${-2.759}_{\times 2}$, ${-2.724}$, ${-2.579}$, ${-2.441}$, ${-2.315}_{\times 2}$, ${-2.168}$, ${-2.158}_{\times 2}$, ${-1.784}$, ${-1.557}$, ${-1.134}$, ${-0.943}_{\times 3}$, ${-0.718}_{\times 2}$, ${-0.625}$, ${-0.577}_{\times 2}$, ${-0.001}$, ${+0.000}_{\times 27}$, ${+0.374}$, ${+0.503}_{\times 2}$, ${+1.347}_{\times 2}$, ${+4.202}_{\times 2}$, ${+4.376}$, ${+5.096}$, ${+5.440}$, ${+5.477}$, ${+6.513}$, ${+6.571}_{\times 2}$, ${+7.522}$, ${+8.770}$, ${+9.008}$, ${+14.928}_{\times 2}$, ${+15.833}$, ${+15.904}$\end{flushleft}\end{minipage}\\[6 pt]{\scriptstyle\begin{minipage}[t]{10cm}\begin{flushleft}$M_{00}\approx0.023832$, $M_{01}\approx0.018667$, $M_{11}\approx0.357987$, $M_{22}\approx-0.277100$, $M_{23}\approx0.007910$, $M_{26}\approx0.016289$, $M_{27}\approx-0.185922$, $M_{33}\approx-0.117652$, $M_{36}\approx0.412728$, $M_{44}{\approx}M_{55}\approx0.049979$, $M_{66}\approx0.107737$, $M_{67}\approx-0.006734$, $M_{77}\approx-0.194761$, $M_{\dot0\dot0}{\approx}M_{\dot1\dot1}\approx0.302520$, $M_{\dot2\dot2}{\approx}M_{\dot3\dot3}\approx0.119245$, $M_{\dot4\dot4}{\approx}M_{\dot5\dot5}\approx-0.138819$, $M_{\dot6\dot6}{\approx}M_{\dot7\dot7}\approx-0.282946$\end{flushleft}\end{minipage}}\\\end{longtable}}


\label{S:sol[-0.178838]1388302}{\small\begin{longtable}{L}\somega{\approx-0.178838}{01388302}: \mathfrak{so}(8)\rightarrow \mathfrak{u}(1),\quad \text{BF-unstable}\\V/g^2\approx-13.8830231447,\quad |\nabla V/g^2|<{7.5}\cdot10^{-12}\\m^2/m_0^2[\psi]:\begin{minipage}[t]{10cm}\begin{flushleft}\scriptsize ${+2.220}_{\times 2}$, ${+2.423}_{\times 2}$, ${+3.551}_{\times 2}$, ${+3.954}_{\times 2}$\end{flushleft}\end{minipage}\\m^2/m_0^2[\chi]:\begin{minipage}[t]{10cm}\begin{flushleft}\scriptsize ${+0.024}_{\times 2}$, ${+0.045}_{\times 2}$, ${+0.083}_{\times 2}$, ${+0.139}_{\times 2}$, ${+0.227}_{\times 2}$, ${+0.272}_{\times 2}$, ${+0.362}_{\times 2}$, ${+0.627}_{\times 2}$, ${+0.929}_{\times 2}$, ${+0.981}_{\times 2}$, ${+1.606}_{\times 2}$, ${+2.880}_{\times 2}$, ${+3.766}_{\times 2}$, ${+4.151}_{\times 2}$, ${+5.035}_{\times 2}$, ${+5.206}_{\times 2}$, ${+6.130}_{\times 2}$, ${+6.530}_{\times 2}$, ${+7.450}_{\times 2}$, ${+7.650}_{\times 2}$, ${+8.881}_{\times 2}$, ${+9.580}_{\times 2}$, ${+9.691}_{\times 2}$, ${+10.059}_{\times 2}$, ${+10.482}_{\times 2}$, ${+13.852}_{\times 2}$, ${+14.204}_{\times 2}$, ${+15.815}_{\times 2}$\end{flushleft}\end{minipage}\\m^2/m_0^2[F]:\begin{minipage}[t]{10cm}\begin{flushleft}\scriptsize ${+0.000}$, ${+0.056}_{\times 2}$, ${+0.130}$, ${+0.171}$, ${+0.415}$, ${+0.443}_{\times 2}$, ${+1.892}$, ${+2.709}_{\times 2}$, ${+2.985}$, ${+3.131}$, ${+3.515}_{\times 2}$, ${+3.673}_{\times 2}$, ${+3.905}$, ${+4.606}$, ${+4.610}$, ${+4.993}$, ${+5.059}$, ${+5.357}_{\times 2}$, ${+7.891}$, ${+8.302}$, ${+8.838}$, ${+9.338}$\end{flushleft}\end{minipage}\\m^2/m_0^2[\phi]:\begin{minipage}[t]{10cm}\begin{flushleft}\scriptsize ${-2.791}$, ${-2.653}_{\times 2}$, ${-2.542}_{\times 2}$, ${-2.489}$, ${-2.420}$, ${-2.266}_{\times 2}$, ${-2.208}$, ${-1.746}$, ${-1.526}$, ${-1.203}$, ${-1.078}_{\times 2}$, ${-1.003}_{\times 2}$, ${-0.950}$, ${-0.803}_{\times 2}$, ${-0.781}$, ${+0.000}_{\times 29}$, ${+0.010}$, ${+0.651}$, ${+1.373}_{\times 2}$, ${+3.374}$, ${+3.577}$, ${+5.125}_{\times 2}$, ${+5.937}_{\times 2}$, ${+5.961}$, ${+6.704}$, ${+7.138}$, ${+7.772}$, ${+8.736}$, ${+8.775}$, ${+11.712}_{\times 2}$, ${+11.882}$, ${+12.064}$\end{flushleft}\end{minipage}\\[6 pt]{\scriptstyle\begin{minipage}[t]{10cm}\begin{flushleft}$M_{00}{\approx}M_{11}\approx-0.011876$, $M_{22}\approx0.206232$, $M_{26}\approx0.091880$, $M_{27}\approx0.392044$, $M_{33}\approx-0.169572$, $M_{36}\approx-0.149145$, $M_{37}\approx0.034954$, $M_{44}\approx0.398687$, $M_{45}\approx-0.082520$, $M_{55}\approx0.010811$, $M_{66}\approx-0.275742$, $M_{67}\approx0.032009$, $M_{77}\approx-0.146664$, $M_{\dot0\dot0}{\approx}M_{\dot1\dot1}\approx0.275041$, $M_{\dot2\dot2}{\approx}M_{\dot3\dot3}\approx0.156221$, $M_{\dot4\dot4}{\approx}M_{\dot5\dot5}\approx-0.172111$, $M_{\dot6\dot6}{\approx}M_{\dot7\dot7}\approx-0.259152$\end{flushleft}\end{minipage}}\\\end{longtable}}


\label{S:sol[-0.178355]1389127}{\small\begin{longtable}{L}\somega{\approx-0.178355}{01389127}: \mathfrak{so}(8)\rightarrow \mathfrak{u}(1),\quad \text{BF-unstable}\\V/g^2\approx-13.8912772896,\quad |\nabla V/g^2|<{5.6}\cdot10^{-13}\\m^2/m_0^2[\psi]:\begin{minipage}[t]{10cm}\begin{flushleft}\scriptsize ${+2.224}_{\times 2}$, ${+2.415}_{\times 2}$, ${+3.549}_{\times 2}$, ${+3.949}_{\times 2}$\end{flushleft}\end{minipage}\\m^2/m_0^2[\chi]:\begin{minipage}[t]{10cm}\begin{flushleft}\scriptsize ${+0.025}_{\times 2}$, ${+0.046}_{\times 2}$, ${+0.083}_{\times 2}$, ${+0.139}_{\times 2}$, ${+0.227}_{\times 2}$, ${+0.271}_{\times 2}$, ${+0.362}_{\times 2}$, ${+0.628}_{\times 2}$, ${+0.930}_{\times 2}$, ${+0.978}_{\times 2}$, ${+1.615}_{\times 2}$, ${+2.881}_{\times 2}$, ${+3.766}_{\times 2}$, ${+4.150}_{\times 2}$, ${+5.038}_{\times 2}$, ${+5.182}_{\times 2}$, ${+6.093}_{\times 2}$, ${+6.525}_{\times 2}$, ${+7.459}_{\times 2}$, ${+7.639}_{\times 2}$, ${+8.895}_{\times 2}$, ${+9.581}_{\times 2}$, ${+9.661}_{\times 2}$, ${+10.043}_{\times 2}$, ${+10.481}_{\times 2}$, ${+13.780}_{\times 2}$, ${+14.196}_{\times 2}$, ${+15.798}_{\times 2}$\end{flushleft}\end{minipage}\\m^2/m_0^2[F]:\begin{minipage}[t]{10cm}\begin{flushleft}\scriptsize ${+0.000}$, ${+0.056}_{\times 2}$, ${+0.131}$, ${+0.172}$, ${+0.416}$, ${+0.447}_{\times 2}$, ${+1.880}$, ${+2.703}_{\times 2}$, ${+3.001}$, ${+3.132}$, ${+3.526}_{\times 2}$, ${+3.675}_{\times 2}$, ${+3.895}$, ${+4.589}$, ${+4.616}$, ${+4.992}$, ${+5.057}$, ${+5.323}_{\times 2}$, ${+7.876}$, ${+8.286}$, ${+8.818}$, ${+9.326}$\end{flushleft}\end{minipage}\\m^2/m_0^2[\phi]:\begin{minipage}[t]{10cm}\begin{flushleft}\scriptsize ${-2.794}$, ${-2.647}_{\times 2}$, ${-2.544}_{\times 2}$, ${-2.485}$, ${-2.420}$, ${-2.263}_{\times 2}$, ${-2.204}$, ${-1.745}$, ${-1.524}$, ${-1.205}$, ${-1.087}_{\times 2}$, ${-1.006}_{\times 2}$, ${-0.950}$, ${-0.808}_{\times 2}$, ${-0.785}$, ${-0.009}_{\times 2}$, ${+0.000}_{\times 28}$, ${+0.660}$, ${+1.371}_{\times 2}$, ${+3.337}$, ${+3.541}$, ${+5.146}_{\times 2}$, ${+5.929}_{\times 2}$, ${+5.976}$, ${+6.732}$, ${+7.153}$, ${+7.787}$, ${+8.728}$, ${+8.779}$, ${+11.654}_{\times 2}$, ${+11.802}$, ${+11.987}$\end{flushleft}\end{minipage}\\[6 pt]{\scriptstyle\begin{minipage}[t]{10cm}\begin{flushleft}$M_{00}{\approx}M_{11}\approx-0.013323$, $M_{22}\approx0.109004$, $M_{23}\approx0.166303$, $M_{26}\approx-0.250471$, $M_{27}\approx0.250202$, $M_{33}\approx-0.069757$, $M_{36}\approx-0.244327$, $M_{37}\approx-0.000523$, $M_{44}\approx-0.006682$, $M_{55}\approx0.416940$, $M_{66}\approx-0.180422$, $M_{67}\approx-0.064825$, $M_{77}\approx-0.242437$, $M_{\dot0\dot0}{\approx}M_{\dot1\dot1}\approx0.274504$, $M_{\dot2\dot2}{\approx}M_{\dot3\dot3}\approx0.156920$, $M_{\dot4\dot4}{\approx}M_{\dot5\dot5}\approx-0.172773$, $M_{\dot6\dot6}{\approx}M_{\dot7\dot7}\approx-0.258651$\end{flushleft}\end{minipage}}\\\end{longtable}}


\label{S:sol[-0.111993]1749289}{\small\begin{longtable}{L}\somega{\approx-0.111993}{01749289}: \mathfrak{so}(8)\rightarrow \mathfrak{so}(3),\quad \text{BF-unstable}\\V/g^2\approx-17.4928962398,\quad |\nabla V/g^2|<{2.2}\cdot10^{-13}\\m^2/m_0^2[\psi]:\begin{minipage}[t]{10cm}\begin{flushleft}\scriptsize ${+2.287}_{\times 4}$, ${+3.592}_{\times 4}$\end{flushleft}\end{minipage}\\m^2/m_0^2[\chi]:\begin{minipage}[t]{10cm}\begin{flushleft}\scriptsize ${+0.189}_{\times 4}$, ${+0.284}_{\times 8}$, ${+0.460}_{\times 4}$, ${+1.054}_{\times 4}$, ${+2.525}_{\times 4}$, ${+4.189}_{\times 8}$, ${+5.156}_{\times 4}$, ${+8.033}_{\times 4}$, ${+8.783}_{\times 4}$, ${+9.149}_{\times 4}$, ${+11.171}_{\times 4}$, ${+14.370}_{\times 4}$\end{flushleft}\end{minipage}\\m^2/m_0^2[F]:\begin{minipage}[t]{10cm}\begin{flushleft}\scriptsize ${+0.000}_{\times 3}$, ${+0.157}$, ${+0.209}$, ${+0.392}$, ${+1.077}_{\times 3}$, ${+2.694}_{\times 3}$, ${+2.877}_{\times 3}$, ${+3.565}$, ${+4.438}$, ${+4.593}_{\times 3}$, ${+4.871}_{\times 3}$, ${+5.296}$, ${+7.041}$, ${+7.245}$, ${+7.719}$, ${+8.717}$\end{flushleft}\end{minipage}\\m^2/m_0^2[\phi]:\begin{minipage}[t]{10cm}\begin{flushleft}\scriptsize ${-2.873}$, ${-2.068}$, ${-1.887}_{\times 3}$, ${-1.704}_{\times 3}$, ${-1.438}_{\times 3}$, ${-1.407}_{\times 5}$, ${-1.155}_{\times 3}$, ${-1.096}$, ${-0.858}$, ${-0.657}_{\times 3}$, ${+0.000}_{\times 28}$, ${+1.041}_{\times 3}$, ${+1.539}$, ${+5.253}$, ${+5.263}$, ${+6.344}_{\times 3}$, ${+7.887}_{\times 3}$, ${+7.954}$, ${+8.125}_{\times 3}$, ${+10.371}$, ${+11.141}$\end{flushleft}\end{minipage}\\[6 pt]{\scriptstyle\begin{minipage}[t]{10cm}\begin{flushleft}$M_{00}\approx0.627034$, $M_{11}\approx0.609774$, $M_{22}\approx-0.088239$, $M_{33}{\approx}M_{44}{\approx}M_{55}\approx-0.171244$, $M_{66}\approx-0.291025$, $M_{77}\approx-0.343813$, $M_{\dot0\dot2}{\approx}-M_{\dot1\dot3}{\approx}-M_{\dot4\dot6}{\approx}-M_{\dot5\dot7}\approx0.200948$, $M_{\dot0\dot5}{\approx}M_{\dot1\dot4}{\approx}M_{\dot2\dot7}{\approx}-M_{\dot3\dot6}\approx-0.086725$\end{flushleft}\end{minipage}}\\\end{longtable}}

\section{Connecting vacua in singular limits}\label{sec:connecting}

\subsection{Gauge group contractions in the embedding tensor formalism}

As we noticed in the previous section, when following the $\omega$-trajectory of a vacuum it can happen that the solution approaches the boundary of $\E7/\SU(8)$ when we move close to some critical value $\omega_*$, so that one or more scalar field vevs blow up in the limit $\omega\to\omega_*$.
This behaviour was already noticed in some analytical computations \cite{DallAgata:2012mfj,Borghese:2012qm,Borghese:2013dja,Gallerati:2014xra}.
For instance, in \cite{DallAgata:2012mfj} four distinct vacua preserving \SO(7) gauge symmetry were found for every $\omega\neq0$, while only three such vacua exist in the original $\omega=0$ theory. 
Following the `missing' vacuum from a non-vanishing deformation parameter towards $\omega=0$, one finds that it moves at infinite distance in scalar field space.

While this behaviour implies that certain vacua disappear from the spectrum of \SOom gauged supergravity at special values of $\omega$, it offers an opportunity to study other gauged theories.
It has been well-established that different gaugings of the same supergravity are connected through singular limits in scalar field space, through a procedure in which the diverging vevs are reabsorned into the embedding tensor (i.e., the gauge couplings) by a duality transformation, and an overall rescaling of the latter is performed to keep it finite \cite{Fischbacher:2003yw,Catino:2013ppa}.%
\footnote{The procedure is analogous to In\"onu--Wigner contractions of a Lie algebra, the difference being that we act with singular duality transformations on $X_{MN}{}^P$ -- thus affecting also the gauge connection -- rather than singular \GL(n,\bbR) transformations on the Lie structure constants.}
For $D=4$ maximal supergravity, this amounts to considering some $U(\lambda)\in\E7$ with $\lambda\in\bbR$, such that $U(\lambda)$ becomes singular when we send $\lambda$ to some limit value $\lambda_*$.
Then, we dress the embedding tensor
\begin{equation}
X_{MN}{}^P 
\to \widehat X(\lambda)_{\,MN}{}^P 
= U(\lambda)_M{}^R U(\lambda)_N{}^S X_{RS}{}^T U(\lambda)^{-1}{}_T{}^P\,.
\end{equation}
Some entries of the rotated embedding tensor $\widehat X(\lambda)_{\,MN}{}^P$ will diverge in the limit $\lambda\to\lambda_*$.
We may then rescale the embedding tensor before taking the limit, by an appropriate power $p$ of $\lambda-\lambda_*$, such that
\begin{equation}\label{limit with rescaling}
\lim_{\lambda\to\lambda_*} 
(\lambda-\lambda_*)^{-p} \widehat X(\lambda)_{\,MN}{}^P
\ =\
\widetilde X_{MN}{}^P\,,\quad\text{finite}
\end{equation}
We take the exponent $p$ that gives a finite but non-vanishing result.
Because the quadratic constraint \eqref{QC1} is \E7 invariant, it is automatically satisfied after the singular limit and $\widetilde X_{MN}{}^P$ (or equivalently $\widetilde\Theta_M{}^\alpha$) defines a consistent gauged supergravity.

Let us now identify $\lambda$ with $\omega$ and let $U(\omega)$ determine the vev of the scalar fields for a chosen vacuum solution, which moves in field space when $\omega$ is changed in some interval $(\omega_0,\,\omega_*)$ and reaches the boundary of $\E7/\SU(8)$ when $\omega\to\omega_*$.
The \SOom embedding tensor is denoted by $X(\omega)_{MN}{}^P$ (without hat) and is finite for any value of the deformation.
We then define 
\begin{equation}\label{X omega}
\widehat X(\omega)_{\,MN}{}^P 
= U(\omega)_M{}^R U(\omega)_N{}^S X(\omega)_{RS}{}^T U(\omega)^{-1}{}_T{}^P\,.
\end{equation}
When $\omega\in(\omega_0,\,\omega_*)$, $\widehat X(\omega)_{\,MN}{}^P $ still defines the \SOom gauging, but now the chosen vacuum has been moved to the origin of scalar field space so that \eqref{vac cond} is satisfied for $\phi=0$ and for any $\omega\in(\omega_0,\,\omega_*)$. 
As a consequence, when we take the limit to $\omega_*$, with appropriate rescaling as in \eqref{limit with rescaling}, the vacuum condition for the new gauging $\widetilde X_{MN}{}^P$ will still be satisfied at $\phi=0$.
This procedure then allows us to construct novel solutions of (potentially) novel gauged supergravities by taking singular limits in $\omega$ space of interesting \SOom vacua.

It is rather straightforward now to see the fate of some simple \SOom vacua when they reach the boundary.
The \SO(7) and $\rmG_2$ vacua disappearing from the \SOom theory for $\omega\to0$ are then recovered in the dyonic \ISO(7) theory, as determined in \cite{Borghese:2012qm}.
A similar fate was noticed for a non-superymmetric AdS vacuum found in \cite{Borghese:2012zs} which drops out of the \SOom theory at a special value of $\omega$ and appears instead in a $\rmU(4)\ltimes\bbR^{12}$ gauging.
We will prove later that there is only one such gauging and that it is part of the `dyonic CSO' family defined in \cite{DallAgata:2011aa}.
Interestingly, the gauged models obtained by such singular limits may admit a geometric uplift to string theory, despite the fact that the \SOom models for $\omega\neq0$ do not.
The dyonic \ISO(7) theory is uplifted to massive IIA supergravity on $S^6$ \cite{Guarino:2015jca,Guarino:2015vca}, and all other `dyonic CSO' gaugings admit uplifts to products of spheres, hyperboloids and tori \cite{Inverso:2016eet}. 
In the next sections we apply numerically the approach described above and identify some  supersymmetric solutions of $\rmU(4)\ltimes\bbR^{12}$ gauged supergravity.

\subsection{From the \texorpdfstring{$\mathrm{SO}(3)\;\mathcal{N}=1$}{SO(3) N=1} vacuum to the dyonic \texorpdfstring{$\rmU(4)\ltimes\bbR^{12}$}{U(4) ⋉ R¹²} gauging}

This section describes a generic, numerical procedure to obtain embedding tensors in singular limits.
We will use the $\SO(3)\;\mathcal{N}=1$ vacuum of the \SOom theories as our example, focussing on the~$\omega/\pi\to1/2$ end of the $\omega$-trajectory shown in figure~\ref{fig:so3n1_flow}.
The three technical challenges that have to be overcome here are that (a) for every point along the trajectory, the gauge group remains the initial one, whereas we are interested in the limit of the (suitably re-scaled) embedding tensor and its associated gauge group, (b) numerical accuracy may limit quite severely how far we can go on a trajectory, at least with TensorFlow, which is currently limited to hardware-supported 64-bit float accuracy, and (c) when starting from an algebraically complicated solution, some ingenuity may be required to arrive at an algebraically simple expression for the limiting gauging.

Let us first address the latter point.
Depending on the algebraic complexity of the starting solution, one may want to make use of the opportunity to absorb only some, but not all scalar vevs into the embedding tensor. 
Loosely speaking, instead of absoring all scalar vevs into the embedding tensor as we did in~\eqref{X omega}, we encode in $U(\omega)$ only those scalar parameters that diverge when $\omega\to\omega_*$ (which ultimately are the ones responsible for the gauge group change).
Other scalar vevs that stay finite in the limit are kept in the $\E7/\SU(8)$ coset representative and will locate in field space the solution that corresponds to the endpoint of the~$\omega$-trajectory in the new gauging.
For example, focussing on the $\SO(3)\;\mathcal{N}=1$ trajectory in figure~\ref{fig:so3n1_flow}, we rescale the embedding tensor in an $\omega$-dependent way such that along the trajectory, we have~$g^{-2}V(\omega)=-6$.
Observing that~$z_3$ approaches the boundary of the Poincare disk while~$z_{1,2}$ stay finite, and using the fact that the seven~$\SL(2,\bbR)/\SO(2)$ subgroups of~$\E7$ that act on the seven~$z_j$ all commute, we retain~$z_{1,2}$ as scalar parameters of the coset representative, but absorb the diverging parameters~($z_3$) to the embedding tensor.

Depending on the equilibrium under study, such a splitting can give us considerable algebraic simplification.
Even when only numerical data are available, convergence acceleration methods may then allow us to extract and confirm the limit gauging.
The full procedure is described below based on the $\SO(3)\;\mathcal{N}=1$ example but immediately generaliseable to any other trajectory.
\begin{enumerate}
  \item Sample points on the~$z_{1,2,3}(\omega)$ trajectories as~$\omega$ approaches
    its limit value, here~$\omega\to\omega_{\rm lim}=\pi/2$, up to~$\omega_{\rm max}$.
  \item Use numerical spline interpolation to obtain numerical estimates for the diverging
    scalar parameters near~$\omega_{\rm max}$ for~$\omega$-values that
    approach~$\omega_{\rm max}$ in geometric progression. (We need about 20 such values.)
  \item Use minimization to refine each of these numerical estimates to an exact location.
    (It here matters that we parametrized the scalars in such a way that we no longer have
     the freedom to move along a~$\SO(8)$-orbit.)
  \item For each such sample, absorb the $z_3$ vev into the embedding tensor as described in equation~\eqref{X omega}, and evaluate the latter to good numerical accuracy. This also needs to be scaled to keep~$g^{-2}V(\Theta)=-6$.
  \item Use iterated (such as: 5-fold iterated) Aitken
    acceleration\footnote{This procedure is nicely explained e.g. in \cite{abelson1996structure}, section~3.5.3.}
    on the numerically-nonzero entries of $\widehat X(\omega)_{MN}{}^P$ to
    estimate its limit $\widetilde{X}_{MN{}^P} = \lim_{\omega\to\omega_*}\widehat X(\omega)_{MN}{}^P$.
  \item For this limit embedding tensor, numerically verify that it satisfies the (linear and quadratic)
    gaugeability conditions. Then, construct from it: (a) the inner product on the gauge group,
    (b) the subspace of `null' directions with respect to this inner product, as well as (c)
    the compact subgroup, (d) the rank of the compact subgroup, and (e) the dimension
    of its commutator group.
  \item The data obtained in the previous step might already suffice to characterize the gauge group.
    If not, proceed to determine roots, pick a subspace of positive roots, determine simple roots,
    and identify the Lie algebra.
  \item The embedding tensor obtained in the steps above typically will embed this gauging into~\E7 in a somewhat complicated way. Proceed to construct a more straightforward embedding from the data obtained in the previous steps.
  \item Numerically scanning for equilibria for the numerically approximate limit gauging
    as well as for the reconstructed gauging, establish that they produce two sets of physically
    equivalent equilibria.
  \item If desired, use numerical minimization to find the \E7 rotation that aligns the
    reconstructed embedding tensor with the numerically estimated one.
\end{enumerate}

This procedure is fully generic and can be automated with an algorithm. While it is in principle applicable to any diverging $\omega$-trajectory, irrespective of the complexity of the underlying equilibrium, it can become challenging in practice to obtain good numerical accuracy for the estimate of the limit embedding tensor. We defer a systematic analysis of the contractions of~\SO(8) that arise as limit gaugings under $\omega$-deformation to subsequent investigations.
For the~$\SO(3)\;\mathcal{N}=1$ solution, executing this procedure\footnote{Details are available in the code accompanying this work} shows that the boundary gauging is (as expected) a~$28$-dimensional gauge group that has 12~null directions and 16 compact ones.
The latter form a rank-4 Lie algebra whose commutator subalgebra is a rank-3 Lie algebra.
This allows us to conclude that the gauge algebra must contain $\mathfrak{u}(4)$ as its reductive factor.
The embedding of this subalgebra into the~$\mathfrak{so}(8)$ algebra is characterized by the branching of its ${\bf 8}_{v, s, c}$ representations, which turns out to be 
\begin{equation}
\mathbf8_v \to \mathbf4_{+1/2} + \bar{\bf4}_{-1/2}\,,\qquad
\mathbf8_c \to \mathbf4_{-1/2} + \bar{\bf4}_{+1/2}\,,\qquad
\mathbf8_s \to \mathbf6_{0} + {\bf1}_{+1} + {\bf1}_{-1}\,.
\end{equation}
Given this decomposition, we look for $\mathfrak{u}(4)$-singlets inside the ${\bf 912}$-representation of~$\mathfrak{e}_{7(7)}$ to which the embedding tensor must belong, since the quadratic constraint implies that the emebdding tensor must be gauge invariant.
There are four singlets which, up to \E7 transformations, can be easily identified with a more general choice of the matrices $\theta_{AB}$ and $\xi^{AB}$ introduced in \eqref{SO8 embtens}:
\begin{equation}\label{u4 four singlets theta xi}
\theta_{AB} = \mathop{\rm diag}(a,a,a,a,a,a,b,b)\,,\qquad
\xi^{AB} = \mathop{\rm diag}(d,d,d,d,d,d,c,c)\,,
\end{equation}
subject to the quadratic constraint $ad-bc=0$.
These define special instances of the class of gaugings defined in \cite{DallAgata:2011aa}.
It was proven in \cite{DallAgata:2014tph} that there is a unique gauging within this class with gauge group including a $\rmU(4)$ reductive factor, namely $[\SO(6)\times\SO(2)]\ltimes\bbR^{12}\simeq\rmU(4)\times\bbR^{12}$, which is obtained for instance by setting $b=d=0$ above and keeping $a\neq0\neq c$.
We conclude that the singular limit of the $\SO(3)\;\mathcal{N}=1$ solution of \SOom gauged supergravity, obtained by following the $\omega$ trajectory in figure~\ref{fig:S1442108_trajectory} until we reach the boundary, gives a vacuum solution of the unique $\rmU(4)\times\bbR^{12}$ gauged maximal supergravity defined in \cite{DallAgata:2011aa}.

As a final validation of the claim about the structure of the limit gauge group, we start from the ~$\rmU(4)\ltimes\bbR^{12}$ embedding tensor defined by~\eqref{u4 four singlets theta xi} with $b=c=0$,
verify that it satisfies the gaugeability conditions and scan
for equilibria. Setting also $a=d=2$, one easily finds~$\mathcal{N}=1$
equilibria at~$g^{-2}V\approx-30.116319$
and~$g^{-2}V\approx-30.331881$ which each are readily re-discovered many
times with numerically identical cosmological constant but rather
different gravitino mass spectra.
The cosmological constants match, up to an overall factor of 4 due to different normalisations, with the one of the \SO(3) invariant $\cN=1$ solution described in \cite{Berman:2021ynm} where the analytic expression as well as the full uplift to IIB supergravity is described,\footnote{The analytic analysis in \cite{Berman:2021ynm} was in fact prompted by the numerical results described in this section.} and with the one of the \rmU(1) invariant $\cN=1$ vacuum recently described in \cite{Bobev:2021rtg}.
The varying gravitino mass spectra we find numerically are easily explained by the presence of an axionic flat direction in these solutions as described in \cite{Berman:2021ynm,Bobev:2021rtg}.
It is very feasible to modify the objective function for numerical
minimization from ``stationarity violation is zero'' to
``stationarity violation is zero, plus the gravitino mass spectrum
matches the spectrum observed for the limit vacuum on the boundary'',
and this indeed finds a vacuum for which all particle masses match
the observed values for the limit vacuum. 
Labelling ${\bf U_4\ldots}$ the vacua of this model, we give the following summary of the properties of the resulting \SO(3) invariant numerical solution, which indeed matches the analytic spectrum provided in \cite{Berman:2021ynm}.
\label{U_4:sol_U4_3011631}{\small\begin{longtable}{L}{\bf U_43011631}: \mathfrak{g}\rightarrow\mathfrak{so}(3), \quad \mathcal{N}={1},\quad \text{BF-stable}\\V/g^2\approx-30.1163185001,\quad |\nabla V/g^2|<{4.1}\cdot10^{-8}\\m^2/m_0^2[\psi]:\begin{minipage}[t]{10cm}\begin{flushleft}\scriptsize ${+1.000}$, ${+4.000}_{\times 4}$, ${+4.556}_{\times 3}$\end{flushleft}\end{minipage}\\m^2/m_0^2[\chi]:\begin{minipage}[t]{10cm}\begin{flushleft}\scriptsize ${+0.000}_{\times 3}$, ${+0.113}$, ${+0.444}_{\times 5}$, ${+1.000}_{\times 3}$, ${+1.940}_{\times 5}$, ${+4.000}_{\times 4}$, ${+4.556}_{\times 3}$, ${+5.726}_{\times 5}$, ${+7.298}_{\times 4}$, ${+7.299}_{\times 2}$, ${+8.301}$, ${+9.889}_{\times 5}$, ${+13.701}_{\times 2}$, ${+13.702}_{\times 4}$, ${+16.000}_{\times 5}$, ${+16.252}$, ${+18.222}_{\times 3}$\end{flushleft}\end{minipage}\\m^2/m_0^2[F]:\begin{minipage}[t]{10cm}\begin{flushleft}\scriptsize ${+0.000}_{\times 31}$, ${+2.000}_{\times 4}$, ${+2.421}_{\times 3}$, ${+3.333}_{\times 5}$, ${+6.000}_{\times 4}$, ${+6.690}_{\times 3}$, ${+10.000}_{\times 6}$\end{flushleft}\end{minipage}\\m^2/m_0^2[\phi]:\begin{minipage}[t]{10cm}\begin{flushleft}\scriptsize ${-2.223}$, ${-2.222}_{\times 5}$, ${-2.000}_{\times 3}$, ${-1.551}$, ${-0.889}_{\times 5}$, ${+0.000}_{\times 28}$, ${+1.333}_{\times 5}$, ${+3.420}$, ${+4.744}_{\times 5}$, ${+8.000}_{\times 6}$, ${+9.182}$, ${+10.000}$, ${+10.221}$, ${+11.034}_{\times 5}$, ${+18.000}$, ${+18.284}$\end{flushleft}\end{minipage}\\[6 pt]
\end{longtable}}

\section{Conclusions}\label{sec:conclusions}

This work contains a first detailed, although likely still incomplete, study of $\omega$-deformed $\SO(8)\;\mathcal{N}=8$ supergravity in four dimensions at the `maximally deformed' $\omega=\pi/8$ point.
In total, we present $390$~critical points, $12$~of them having residual supersymmetry, and six more having perturbative stability without residual supersymmetry.
Many of these were not known before.
Additionally, we studied how vacuum solutions move in field space, branch and merge when $\omega$ is changed. 
We then explained in detail the procedure for obtaining new gaugings of four-dimensional supergravity with guaranteed equilibria via a limiting procedure that investigates the behavior of $\omega$-deformation trajectories of vacua as they approach the boundary of the scalar field manifold.
While $\omega$-deformed \SO(8) gauged maximal supergravity does not admit an uplift to 10 or 11 dimensional supergravity \cite{deWit:2013ija,Lee:2015xga}, the gaugings obtained through singular limits may have a higher dimensional origin.
We exemplified this situation by identifying a supersymmetric solution of $\rmU(4)\ltimes\bbR^{12}$ gauged supergravity, whose analytic expression and uplift to type IIB supergravity has been recently carried out by three of the authors \cite{Berman:2021ynm}.
Certainly, the techniques developed here can be exploited much further, to chart the landscape of vacua of \SOom gauged supergravities and study their web of interconnections, as well as to construct novel gaugings with interesting vacuum solutions via singular limits.
Combined with known general techniques to identify the uplift manifolds and ans\"atze of gauged maximal supergravities \cite{Inverso:2017lrz}, these developments will make it possible to identify many new interesting solutions of supergravity and string theory.
A more systematic analysis of the $\omega$-trajectories of \SO(8) gauged maximal supergravity vacua is in developement \cite{TOAPPEAR}, and we expect to come back to these and related ideas in the future.

\section*{Acknowledgments}

This project has received funding from the European Union’s Horizon 2020 research and innovation programme under the Marie Skłodowska-Curie grant agreement No 842991.
DSB gratefully acknowledges the support by Pierre Andurand.

\appendix

\section{Appendix: All solutions}\label{app:allsolutions}

\input{so8c_detailed.tex}

\providecommand{\href}[2]{#2}\begingroup\raggedright\endgroup

\end{document}